\documentclass[%
amsmath,amssymb,
 reprint,%
aip,
cha
]{revtex4-1}

\usepackage{graphicx}
\usepackage[dvipsnames]{xcolor}
\graphicspath{{fig/}} 
\usepackage{dcolumn}
\usepackage{bm}

\usepackage[english]{babel}
\usepackage{lipsum}
\usepackage[]{longtable}

\begin{document}

\title[]{Intervality and coherence in complex networks\\}
\vspace{0.5cm}
\author{Virginia Dom{\'\i}nguez-Garc{\'\i}a} \affiliation{Departamento
  de Electromagnetismo y F{\'\i}sica de la Materia and Instituto
  Carlos I de F{\'\i}sica Te\'orica y Computacional. Universidad de
  Granada.  E-18071, Granada, Spain.}  \author{Samuel Johnson}%
\affiliation{Warwick Mathematics Institute, and Centre for Complexity
  Science, University of Warwick, Coventry, CV4 7AL, United
  Kingdom.}
\author{Miguel A. Mu\~noz} \affiliation{Departamento de
  Electromagnetismo y F{\'\i}sica de la Materia and Instituto Carlos I
  de F{\'\i}sica Te\'orica y Computacional. Universidad de Granada.
  E-18071, Granada, Spain.}

\email{mamunoz@onsager.ugr.es}


\begin{abstract}  
  Food webs -- networks of predators and prey -- have long been known
  to exhibit ``intervality'': species can generally be ordered along a
  single axis in such a way that the prey of any given predator tend
  to lie on unbroken compact intervals. Although the meaning of this
  axis -- identified with a ``niche'' dimension -- has remained a
  mystery, it is assumed to lie at the basis of the highly non-trivial
  structure of food webs. With this in mind, most trophic network
  modelling has for decades been based on assigning species a niche
  value by hand. However, we argue here that intervality should not be
  considered the cause but rather a consequence of food-web
  structure. First, analysing a set of $46$ empirical food webs, we
  find that they also exhibit {\it predator} intervality: the
  predators of any given species are as likely to be contiguous as the
  prey are, but in a different ordering.  Furthermore, this property
  is not exclusive of trophic networks: several networks of genes,
  neurons, metabolites, cellular machines, airports, and words are found to be
  approximately as interval as food webs.  We go on to show that a
  simple model of food-web assembly which does not make use of a niche
  axis can nevertheless generate significant intervality.  Therefore,
  the niche dimension (in the sense used for food-web modelling) could
  in fact be the consequence of other, more fundamental structural
  traits, such as trophic coherence.  We conclude that a new approach
  to food-web modelling is required for a deeper understanding of
  ecosystem assembly, structure and function, and propose that certain
  topological features thought to be specific of food webs are in fact
  common to many complex networks.
\end{abstract}
\maketitle

\begin{quotation} 
  For decades food-web modelling has been based on the idea of a
  ``niche dimension'', according to which the species in an ecosystem
  are considered to be arranged in a specific order, which is
  tantamount to the existence of a one-dimensional hidden dimension.
  This assumption is justified by the empirical observation of a
  topological feature, exhibited by many food webs to a significant
  degree, called ``intervality''.  We show here that intervality is
  not necessarily the hallmark of a hidden niche dimension, but may
  ensue from other food-web structural properties, such as trophic
  coherence.  In fact, we find instances of networks of genes,
  neurons, metabolites, cellular machines, airports, and words which exhibit intervality
  as significant as that of food webs.  These results support a new
  approach to food-web modelling, and suggest that certain features of
  trophic networks are relevant for directed networks in general.
\end{quotation}

\section{Introduction}
\noindent
Charles Darwin concluded {\it On the Origin of Species} reminiscing on
his famous entangled bank, ``clothed with many plants of many kinds,
with birds singing on the bushes, with various insects flitting about,
and with worms crawling through the damp earth, [...] so different
from each other, and dependent on each other in so complex a
manner''.\cite{Darwin} Charles Elton later developed the concept of a
food web -- a network of predators and prey -- as a description for a
community of species,\cite{Elton} and in the eighties such systems
were among the first to be explicitly modelled as random graphs with
specific constraints. \cite{Cohen_1,Cohen_book} With the advent of
ever better ecological data and the explosion of research on complex
networks, much work has gone into analysing and modelling the
structure of food webs, and its relation to population dynamics and
ecosystem
function.\cite{Pimm,Drossel,Dunne_rev,Jordi_book,guimera_foodweb,sole_2001}
Not least among the motivations for such research has been an
awareness that the sixth mass extinction is under way, and that we
must strive to understand ecosystems if we are to protect
them.\cite{de2015estimating}

Field ecologists apply a variety of techniques to infer the predation
links which exist between (and sometimes within) the dozens, or
hundreds, of species making up specific ecosystems. The results of
such observations are sets of trophic networks, or food webs, which can
now be analysed quantitatively, as we go on to do here.  A food web
with $S$ species can be encoded in an $S\times S$ \emph{adjacency
  matrix} $A$, such that the element $A_{ij}$ is equal to one if
species $i$ (the predator) consumes species $j$ (its prey), and zero
if not.  In other words, a food web can be regarded as an unweighted,
directed network in which the nodes are species and the directed edges
represent predation.

\begin{figure}[t]s \begin{center}
\includegraphics[scale=0.6]{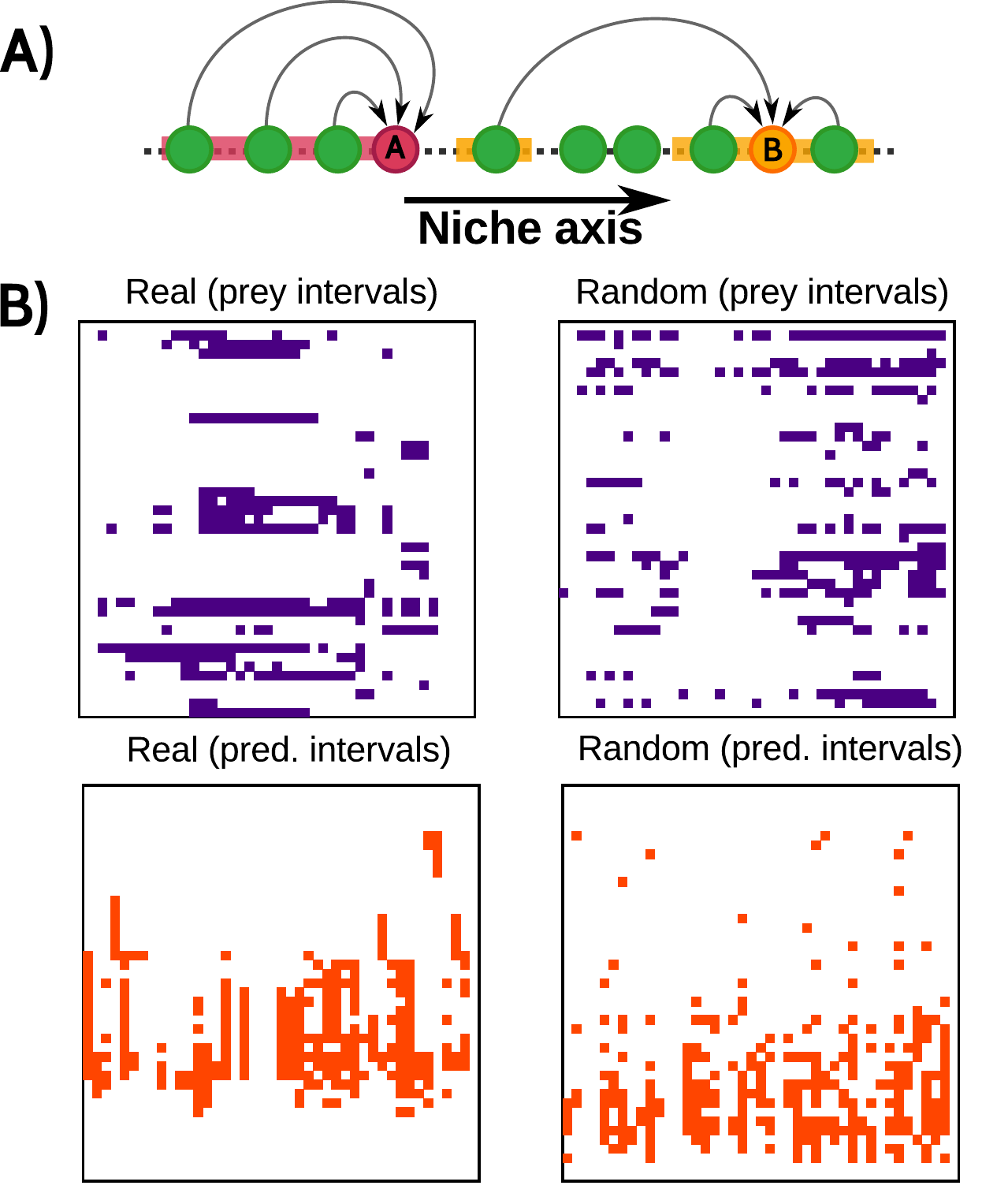} 
\end{center} \caption{A) Examples of interval and non-interval
  ``diets''. Once species have been arranged along some hidden (niche)
  axis, species A is said to have an interval diet if all their prey
  occupy contiguous positions, while species B has a non-interval
  diet. Different orderings may result in different values of diet
  intervality for a given predator species.  The niche ordering is defined as
  the one maximizing the overall level of intervality.  B) Top left
  panel: Adjacency matrix corresponding to the empirically obtained
  food web of Mondego Estuary. \cite{mondego} Filled squares stand for
  predators (vertical axis) consuming the corresponding prey (horizontal
  axis); as usual, the same ordering is used for both rows and
  columns. An ordering of the species has been sought which maximizes
  prey-intervality, $\xi$ (see main text for a definition): observe
  that the prey of any predator tend to be contiguous/interval. Top
  right panel: The same procedure is applied to a randomised version
  of the network which preserves both in- and out-degree sequences
for each node in 
the Mondego food web; observe the
  strong reduction in the level of
  intervality.
  Bottom left panel: The same adjacency as in the top left panel, but this time 
  according to an ordering or elements which maximises predator-intervality, $\eta$
  (see main text). Note that this ordering is clearly different from the one found in the panel above.
  Bottom right panel: Reorganised matrix maximising $\eta$ for a network randomisation.
  } \label{fig_real} \end{figure}

When Joel Cohen first examined a set of such food webs in the
seventies, he discovered that they exhibited a topological property
which he named {\it intervality}: the species could be ordered in a
line in such a way that the prey of any given predator would form a
compact interval.\cite{Cohen,Cohen_book} In terms of the adjacency
matrix, this meant that the columns could be ordered so that elements
would form unbroken horizontal blocks (see Figure \ref{fig_real}A).
This observation re-invigorated the use of an important and old
concept in ecology: the {\it niche}. The term was originally used
simply to refer to a species' habitat \cite{Grinnell} or ecological
role, \cite{Elton} but was then defined by Hutchinson as a ``position
in a multi-dimensional hyperspace'' -- each dimension being some
biologically relevant magnitude. \cite{Hutchinson} The observed
intervality of food webs suggested that predators consume every
species within a particular compact hypervolume of niche space -- and,
moreover, that niche space was (at least to a good approximation)
one-dimensional. \cite{Williams_niche_model,Dunne_rev}

Motivated by the belief that complex systems could come about from
simple rules, several models were put forward to explain the
non-trivial structure of trophic networks.  The cascade model was the
first such attempt.\cite{Cohen_1} The key idea behind this model is
that food webs reflect some inherent hierarchy in which certain
species are above others.  Hence, species are organised on a
hierarchical (niche) axis and are constrained to consume only prey
which are below them. \cite{Cohen} The approach is loosely based on
the fact that predators tend to be larger than their prey, at least in
certain kinds of ecosystem, and so the hierarchy could be regarded as
one defined by body size.  The cascade model was better at reproducing
food-web structure than a fully random graph, but there were features
-- intervality, in particular -- which it could not account for. Then
Williams and Martinez put forward the well-known \emph{niche
model},\cite{Williams_niche_model} in which species are again ordered
along a hierarchical or niche axis, but species are allowed to select
prey now only on a contiguous interval below them (not a random
selection, as in the cascade model). Thus, the niche model has
intervality as a built-in property. It also proved quite successful in
explaining certain other food-webs properties, and, thanks to its
simplicity, it is often taken as the reference model for generating
synthetic food webs.\cite{nicheforever,drosselniche} The popularity
enjoyed by the niche model has served to reinforce the belief that
something akin to its niche axis does in fact form the backbone of
real ecosystems.\cite{Stouffer_robust,Zook}

As progressively better collections of food-web data were gathered, it
became apparent that most trophic networks were not perfectly interval (see
Figure \ref{fig_real}B). Measures of local ``frustration'' were
proposed to capture the distance from perfect
intervality,\cite{Cattin} and more recently the degree of
intervality of food webs has been measured with several continuous
quantities that take values close to unity if most of the prey lie on
unbroken intervals, or approach zero when very few
do.\cite{Stouffer_robust,Zook} A simple way is to measure, for each
predator, the size of its largest unbroken interval as a fraction of
its total number of prey, and use the average value over predators as
a measure of the intervality -- also sometimes called {\it contiguity}
-- $\xi$ of the food web; however, since intervality is defined for a
given ordering of species, some optimization method is required to
find the most interval global ordering, much as is done for magnitudes
such as modularity\cite{Newman_compartments} or
nestedness\cite{plos_nestedness} (see Methods for a more
precise definition).  Several modifications of the niche model have
been proposed to overcome the drawback of perfect intervality that the
niche model introduces by construction. Both the ``generalized niche''
model\cite{Stouffer_GNM} and the ``minimum potential niche''
model\cite{allesina_model} relax the assumption that the prey of a
given predator must form a perfectly continuous interval.  The
rationale is that the imperfect intervality of food webs might be a
sign that there are, in fact, more than one niche
dimension.\cite{allesina_model} Despite these developments, some kind
of niche axis is still generally thought to underlie food-web
structure.

Here, we challenge this view by appealing to three
observations. 

First, we highlight that {\it predator intervality} (the
extent to which the predators of a given species can be arranged on
unbroken intervals;\cite{Zook} i.e. the predator intervality of a
matrix $A$ is the prey intervality of its transpose, $A^T$) seems to
be as general and non-trivial a feature of food webs as {\it
  prey-intervality}.  While it is conceivable that predators might
``choose'' their prey according to a niche axis, how can prey
simultaneously select their predators according to a different
ordering?  

Second, we show that intervality is not a feature peculiar
to food webs, as has usually been assumed: complex networks of various
kinds -- including those of genes, neurons, metabolites, cellular
machines, airports, and words -- exhibit levels of both prey
(column) and predator (row) intervality similar to those of food webs.
Yet it appears far-fetched to suggest that such diverse systems all
owe their structure to some kind of hidden niche axis.  

Third, we show that the recently proposed ``preferential preying''
model of food-web assembly, which does not involve a niche axis,
generates significant intervality.\cite{coherence_PNAS} This is the
first model to correctly reproduce the trophic coherence of food webs
--i.e. the fact that species can be assigned trophic levels and
predators have a tendency to prey upon subsets of prey which are on similar such
levels-- and we find that a degree of intervality is a by-product of
this topological feature.  We go on to propose a version of the
preferential preying model, amended to take account of phylogenetic
constraints, which generates realistic values of intervality without
fitting additional parameter values.

Taken together, we believe these observations call into question the
concept of a niche dimension, whether as an operationally useful
construct for food-web modelling, or as a reality to be uncovered in
nature.  We conclude by discussing what these findings might mean for
our understanding of food-webs and other complex networks.

\section{Results}

\subsection{Intervality}

\noindent
Let us consider a directed network with $S$ nodes and $L$ edges,
defined by the $S\times S$ adjacency matrix $A$.  As we have said, in
the case of a trophic network the nodes are species and the edges
represent predation.  The in- and out-degrees of node $i$ are
$k_i^{in}=\sum_j A_{ij}$ and $k_i^{out}=\sum_j A_{ji}$, and correspond
to the numbers of prey and predators of species $i$, respectively; and
the mean degree is $\langle k\rangle=L/S$.  Basal species, or
autotrophs, have no in-coming edges, and are thus represented by nodes
with $k^{in}=0$. We shall denote with $B$ the number of basal nodes in
a given network.

The top left panel of Fig.\ref{fig_real}.B shows the adjacency matrix
for the food web of Mondego Estuary,\cite{mondego} on the Atlantic coast of Portugal,
with species ordered so as to maximize intervality $\xi$ (see
Methods). Since there are $S!$ possible orderings of the
columns it is not feasible in general to perform an exhaustive search
for the most interval one. Therefore, we proceed as Stouffer {\it et
  al.} \cite{Stouffer_robust} and use a simulated annealing (SA)
algorithm -- described in the Methods section.  For comparison, in the
top right panel of Fig.{\ref{fig_real}.B} we show the best ordering
for a randomisation of the network which preserves
$k^{in}$ and $k^{out}$ for each node.
These plots readily reveal that the empirical
food web is remarkably more interval than its randomised
counterpart. Observe, however, that the random network exhibits a
non-zero level of intervality, since this magnitude is defined for the
most interval ordering out of a great many possible choices.  It is
therefore always necessary to compare empirical values of intervality
to the corresponding random expectations in a null model, in order to
determine the significance of this measurement.

In the bottom left panel of Fig.{\ref{fig_real}.B} we show the
same Mondego Estuary adjacency matrix, but this time an ordering has been
found which maximises predator-intervality, $\eta$: the extent to which
the predators of a given prey species are contiguous \cite{Zook} (as
above, the same procedure has been applied to an ensemble of network
randomisations). The real food web is significantly more
predator-interval, while the random graph has some degree of
predator-intervality (there is a preponderance of vertical
intervals). Interestingly, the ordering of maximum
predator-intervality is different from the one yielding the highest
prey-intervality, as is obvious to the naked-eye upon inspection of
the two patterns. In fact, we shall see that predator-intervality
seems to be as general and non-trivial a feature of food webs as the
oft-cited intervality $\xi$ -- which we shall henceforth refer to as
prey-intervality, to distinguish the two concepts.

\begin{figure}[b]
\begin{center}
\includegraphics[scale=0.27]{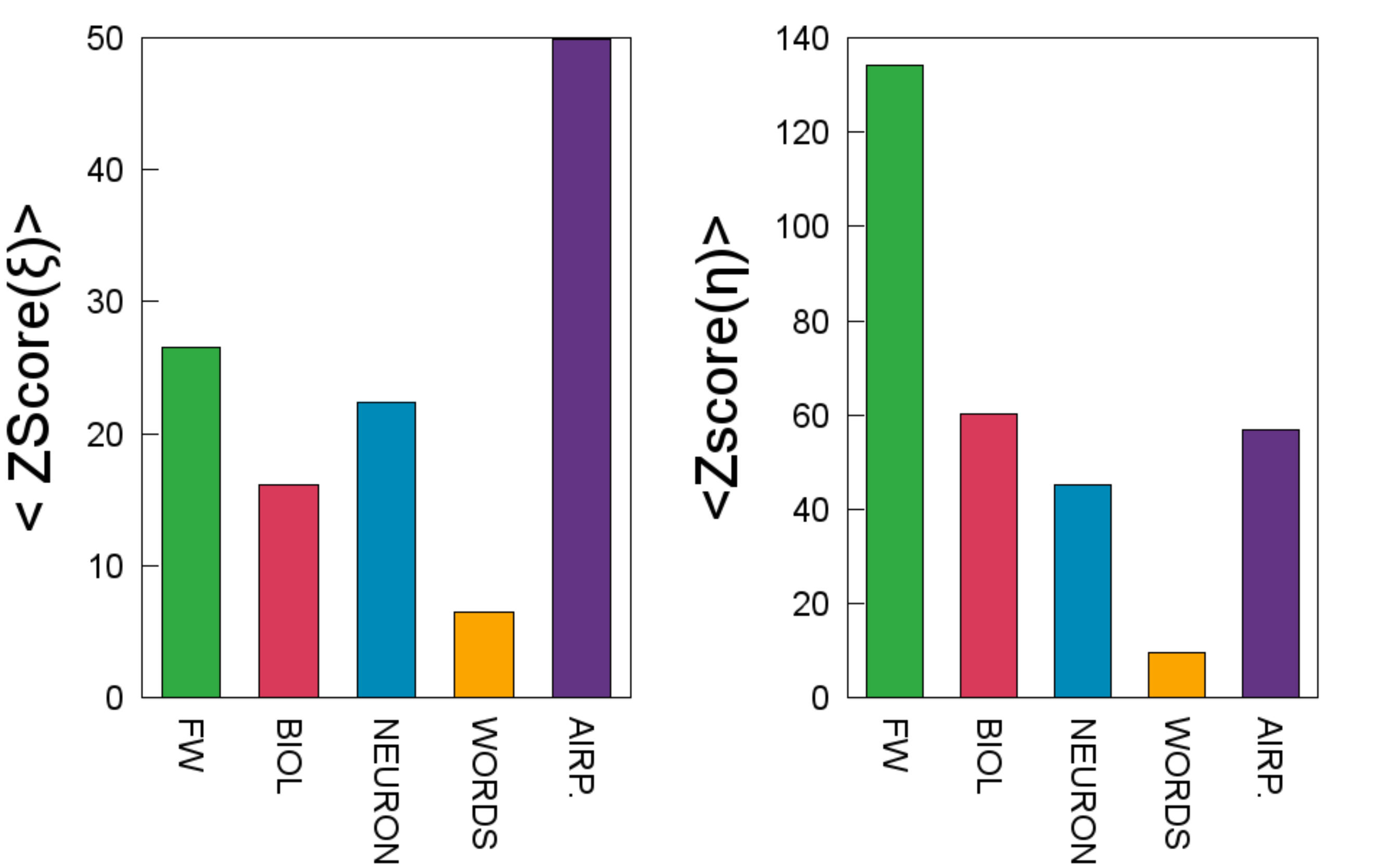}
\end{center}
\caption{Average z-score for prey ($\xi$) and predator ($\eta$)
  intervalities for different type of networks (with respect to a null model
  consisting in randomisations of each network that preserve 
 both $k_{in}$ and $k_{out}$ for each node). Empirical values of the z-score have been
  averaged over all networks in each category. It is evident that
  empirical values lie well outside the random expectation, with high values of z-scores in all cases. 
 Intervality is therefore a non-trivial feature of many kinds of complex network.}
\label{fig_randomisations}
\end{figure}

We now extend this analysis to the entire database of directed
networks we are going to use throughout this work (listed in Table \ref{table_zscore}),
which includes food webs, gene
transcription networks, metabolic networks, networks of cellular signalling, the neural
network of {\it C. elegans}, a word-adjacency network, and a US
airport recommendation network. The two panels of
Fig.{\ref{fig_randomisations}} show average prey-intervality $\xi$,
and average predator-intervality $\eta$, for all networks in the
database, compared to their randomisations. The first observation is
that, as we can see in Fig.\ref{fig_real} for Mondego Estuary, and more generally in Table \ref{table_zscore},
prey-intervality is indeed a common feature of food webs.  In all
cases the empirical values lie several standard deviations from the
random expectations (z-score values greater than $3$ in all but $4$
cases out of $46$ food webs).  Similarly, food webs also exhibit
significant levels of predator-intervality $\eta$. In all cases, the
orderings which maximise prey- and predator-intervality are different.
What is perhaps more interesting is that intervality does not seem to
be limited to food webs; indeed, almost all the other networks in
our database show a similar trend, with levels of both intervalities
similar to those observed in food webs.

Since the original observation that food webs tended to be interval,
the existence of a hidden ``niche dimension'' has been assumed to be
at the root of this deviation from randomness.  However, this
explanation fits ill with the fact that food webs and other kinds of
network exhibit both prey- and predator- intervality to similar degrees
of significance.  It would seem, rather, that a more general reason
must exist which can account for this topological feature.

\subsection{Trophic coherence}
\noindent
One of the most striking characteristics of food webs -- perhaps the
first that springs to mind upon contemplating even a child's drawing
of an ecosystem -- is the existence of a trophic structure. At the
base there are plants, which are consumed by herbivores, which in turn
might be preyed upon by omnivores or primary carnivores, and so the
biomass flows from producers all the way up to top predators.
Ecologists quantify the position of a species in a food web with its
trophic level $l$, such that basal species have $l=1$, herbivores
$l=2$, etc.  More generally --following Levine \cite{Levine_levels}--
one can define the (non-integer) trophic levels in a recursive,
self-consistent way: the trophic level of a given species is equal to
the average level of its prey plus one unit.  That is,
\begin{equation} 
l_i=\frac{1}{k_i^{in}}\sum_j A_{ij}l_j+1
\label{eq_levels}
\end{equation}
if $k_i^{in}>0$, or $l_i=1$ if $k_i^{in}=0$.  Observe that this
definition is easily generalizable to other types of directed network,
and each node can be assigned a trophic level by simply solving the
set of linear equations given by Eq. (\ref{eq_levels}).\cite{loopines}
The only requirements for all nodes to have a well-defined level are
that there should exist at least one basal node ($B>0$), and that
every non-basal node must be on at least one path including a basal
node.\cite{Johnson_spectra} In fact, this definition of trophic level
is similar to other measures of centrality, such as PageRank, with the
difference that the ``$+1$'' term establishes a natural
hierarchy.\cite{dominguez2015ranking}

We have recently shown that the trophic structure of a directed
network can be characterised by a degree of order we call {\it trophic
  coherence}.\cite{coherence_PNAS} For this we define a
variable $x$ for each edge, as $x_{ij}=l_i-l_j$, and consider the
distribution of $x$ over the $L$ edges of a network. The mean is
$\langle x\rangle=1$ by definition, and the homogeneity of the
distribution is the trophic coherence of the network. We can therefore
quantify this feature simply with the standard deviation of $p(x)$,
which we refer to as an {\it incoherence parameter}: $q=\sqrt{\langle
  x^2\rangle-1}$. A highly coherent network ($q\simeq 0$) is one in
which the nodes fall into clear (almost integer) trophic levels, while
a more random system is less coherent ($q>0$).  We have shown that
trophic coherence is key to the linear stability of food webs and,
since it can invert the usually positive relationship between
diversity and stability, might be the solution to Robert May's famous
paradox.\cite{coherence_PNAS,May,McCann} Trophic coherence has
subsequently been found to play an important role in directed networks
of many kinds, including those of neurons, genes, metabolites,
cellular signalling, words, P2P, trade and transportation, and this
feature is intimately related to cycles and feedback loops, graph
eigenspectra and the ubiquity of `qualitatively stable'
systems.\cite{loopines,Johnson_spectra} Elsewhere in this issue,
Klaise \& Johnson show that trophic coherence also 
determines the extent and duration of spreading processes such as 
epidemics or neuronal cascades.\cite{Janis_chaos}
\begin{figure}[b]
\begin{center} 
\includegraphics[scale=0.22]{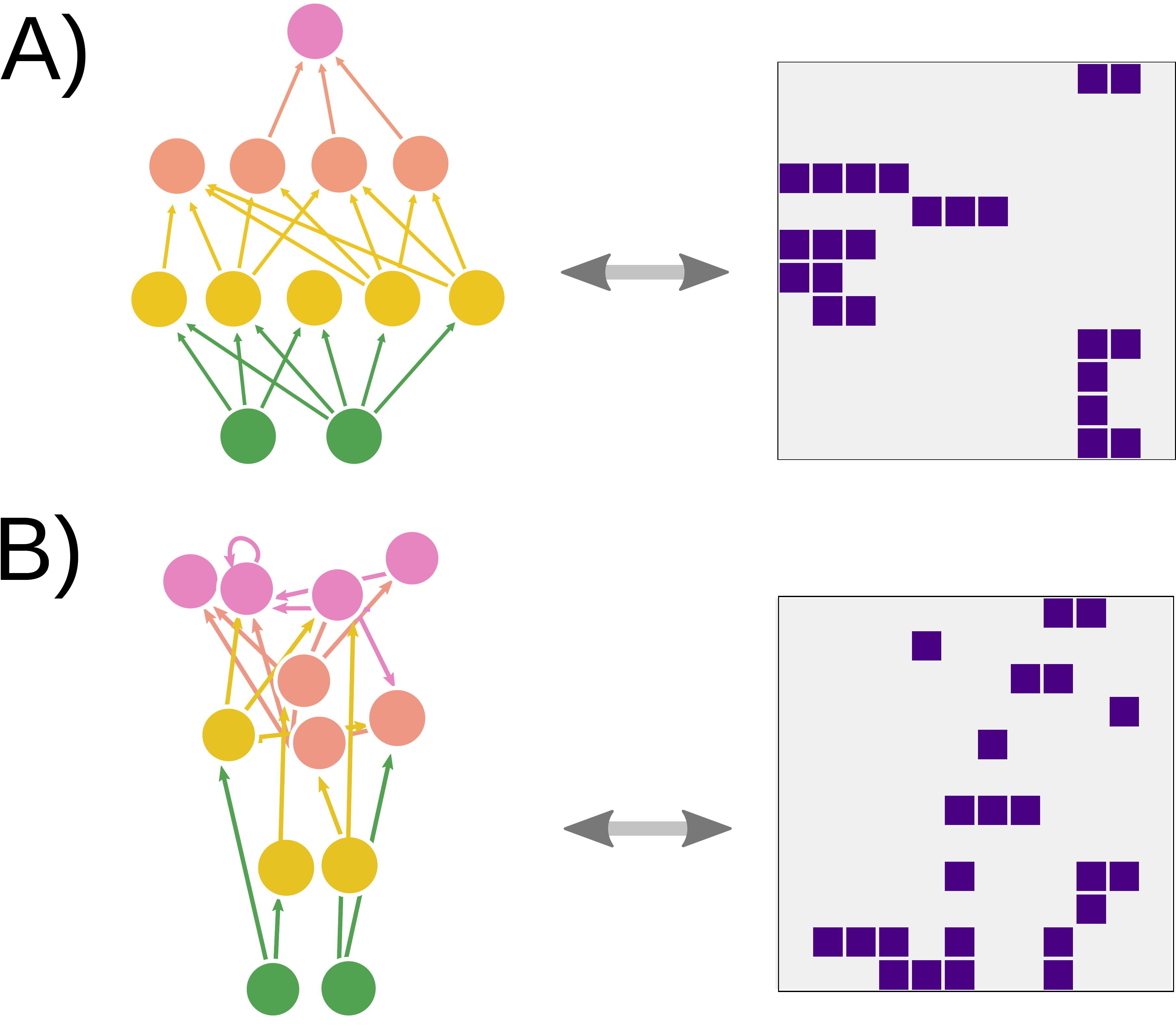}
\end{center}
\caption{A: Network generated with the preferential preying model
  (PPM) with $S = 12$, $B = 2$ and $T = 0$ (the vertical position of
  each node reflects its trophic level) and its corresponding
  adjacency matrix, with rows and columns ordered to maximize
  prey-intervality ($\xi=1$).  B: As A but for a similar randomised network, 
  leading to a smaller intervality $\xi=0.90$.}
\label{fig_model}
\end{figure}

Networks with tunable trophic coherence can be generated with the
recently proposed {\it preferential preying model}
(PPM),\cite{coherence_PNAS} which works as follows.  We begin with $B$
initial nodes (basal species) and no edges. New nodes (consumer species) are added
sequentially to the system until a total of $S$ nodes is
reached.  When a node enters the system its in-neighbours (prey) are
awarded from among available nodes (those already in the network) in
the following way: the first prey species is chosen randomly, and the
rest are chosen with a probability that decays exponentially with the
absolute trophic distance to their initial prey (i.e. with the
absolute difference of trophic levels between its first prey and the
subsequent ones). This probability is set by a parameter $T$ that
determines the degree of trophic specialization of consumers, and
normalised so as to produce an expected number of edges $L$.
The lower the value of $T$ (while $T>0$), the more the network will have a coherent
trophic structure ($q \sim 0$). From the perspective of evolutionary
ecology, this simply means that if a given predator is good at
consuming species X, then its other prey are likely to have similar
trophic levels to that of X, as is observed in nature.  \footnote{This is the
model we use to generate coherent networks in this work; however, we
note that Klaise \& Johnson\cite{Janis_chaos} propose a slightly
different version of the PPM, the main difference being that at high
$T$ their model limits in random graphs instead of acyclic cascade
model networks.}

\begin{figure*}[]
\begin{center}
\includegraphics[width=\textwidth]{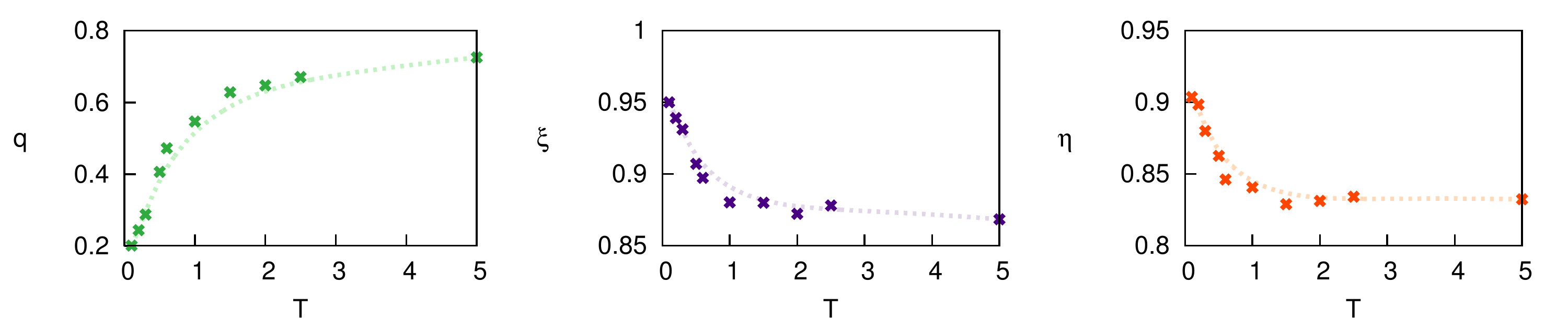}
\end{center}
\caption{Left panel: Incoherence  parameter $q$ as a function of the
  tunable parameter $T$ as obtained in the {\it preferential preying model} (PPM)\cite{coherence_PNAS} ($N=31$ $L=68$ as in the Chesapeake bay food web
 \cite{chesapeake1,chesapeake2}). For each $T$, the reported values
  (crosses) are averages over $100$ independent network realisations.
  As $T$ increases the generated networks are more incoherent. 
  The Central (Right) panel shows prey (predator) intervality $\xi$ ($\eta$), as measured in the same networks. 
  Discontinuous lines are just interpolations.}
\label{fig_q_inter}
\end{figure*}

Food webs tend to be very significantly coherent, and this is key to
their stability and other properties. However, niche-based models are
not able to reproduce this feature, and generate networks which are
barely more coherent than random graphs.\cite{coherence_PNAS} This
shows that intervality is not a sufficient condition for trophic
coherence.  Is it possible, though, that trophic coherence induces
intervality?  Figure \ref{fig_model}A displays an example of a
maximally coherent network ($q=0$), obtained with a low $T$, while
figure \ref{fig_model}B is an instance of a less coherent network
($q>0$), obtained with a high $T$ (see also Fig.\ref{fig_q_inter}). 
Beside each of them appears its
adjacency matrix, where species have been ordered in such a way as to
maximise prey intervality, $\xi$. The maximally coherent network is
also perfectly interval, while the incoherent one has gaps which
cannot be eliminated by changing the arrangement of nodes in any
way. This trend is more clearly illustrated in Fig.4,
which shows that both prey- and predator-intervalities decrease
monotonically with $T$.

For an intuitive understanding of why intervality emerges in the
presence of trophic coherence, imagine a simple food web consisting
only of $n_a$ predators that prey upon a different set of $n_b$
species. It will be more likely that there is an ordering of the prey
such that $\xi$ is high --or one of predators such that $\eta$ is-- if
$n_a$ and $n_b$ are small (i.e. if the adjacency matrix contains few
rows and columns). If a food web is perfectly stratified ($q=0$) so
that species on level $l$ are only consumed by those on level $l+1$,
then the network can be seen as a superposition of many of these
(independent) simple situations ($A$ will have a ``modular'' or block
structure). Given that at each pair of levels the number of species is
significantly smaller than the total number $S$, a global ordering can
be found yielding as good a $\xi$ as in the simple example -- or a
different ordering for as good an $\eta$. Networks that are thus
highly coherent will have higher values of both intervalities than
those lacking this structure. 

\subsection{Topological features contributing to Intervality }
\noindent
The reasoning described above suggests that trophic coherence should
not be the only feature related to intervality. Indeed, as
illustrated below, trophic coherence can only account for a fraction
of the total intervality observed in empirical networks.

In general, any property which had the effect of creating modularity
--i.e. de-coupling certain non-zero matrix elements from others, in
the sense that their rows or columns could be shuffled without
affecting the ordering of other elements-- would be conducive to
higher levels of intervality. In other words, we should expect a high
prey-intervality in networks in which nodes sharing any in-neighbours
tended to share a high proportion of them; and the same goes for
predator-intervality and shared out-neighbours.  To capture this
feature, we can define \emph{in-complementarity} $c_{in}$ as the mean
number of shared in-neighbours over all pairs of nodes with any shared
in-neighbours; and, similarly, \emph{out-complementarity} $c_{out}$ as
the mean number of shared out-neighbours over all pairs of nodes with
any shared out-neighbours.

We have measured the relationships between intervality and several
topological properties (including $S$, $\langle k\rangle$, $q$ and
complementarities $c_{in}$ and $c_{out}$). The correlation between
$\xi$ and $c_{in}$ has a Pearson coefficient of $r=0.83$, precisely
the same as that of $\eta$ and $c_{out}$. In other words,
complementarity --i.e. the proportions of shared in-neighbours and of
shared out-neighbours-- accounts for approximately $70\%$ of the
variance in prey- and predator-intervalities, respectively, across a
broad range of empirical networks.  The correlations with trophic
coherence itself are lower but still significant ($r \approx 0.55$).
The mean degree, $\langle k\rangle$, and the number of nodes, $S$, are
both negatively correlated with both kinds of intervality ($r \approx
-0.67$ and $r \approx -0.45$, respectively), as we would expect from
the reasoning above.

\subsection{Modelling networks with coherence and intervality}
\noindent
Most past attempts to explain food-web intervality have focused on
identifying the biological factors which might yield a niche
dimension, such as body size.\cite{Stouffer_robust,Zook,arenas_model} However, the
work by Cattin {\it et al.}\cite{Cattin} is different in that it
considers the effects of phylogeny, albeit within the framework of a
niche-based model. Crucially, the existence of phylogenetic
relationships between species -- i.e. of common ancestry -- was not
modelled as a niche axis, but as a ``diet overlap'' for certain pairs
of species.  The thinking is that species which are closely related to
each other (phylogenetically similar) will tend to share many of their
prey.  Cattin and colleagues found that phylogenetic constraints,
modelled in this way, increase intervality\cite{Cattin} -- and
subsequent empirical work has reinforced this
connection.\cite{eklof_philo,pulin_philo} These results are in keeping
with the view that any mechanism with the effect of making nodes
topologically similar (such as close relatives having many shared prey
or predators, in the case of food webs) will also lead to seemingly
non-random levels of intervality.  But since topologically similar
nodes can exist for many different reasons, it is not just food webs
which exhibit intervality, but directed networks of many different
kinds.

In the case of food webs, we know that both trophic coherence and a
phylogenetic signal are present. To explore whether these properties
might be sufficient to explain the observed intervality, we extend the
PPM to account for the effect of phylogenetic relations in a simple
parameter-free way.  For this we draw inspiration from the ``nested
hierarchy'' model of Cattin {\it et al.},\cite{Cattin} and define a
version of the PPM in which new nodes are assigned in-neighbours
(prey) according not only to their trophic levels, as before, but also
with a preference for nodes which already share out-neighbours
(predators) whenever possible (see Methods for a detailed description
of this model).
\begin{figure*}[]
\begin{center}
\includegraphics[scale=0.5]{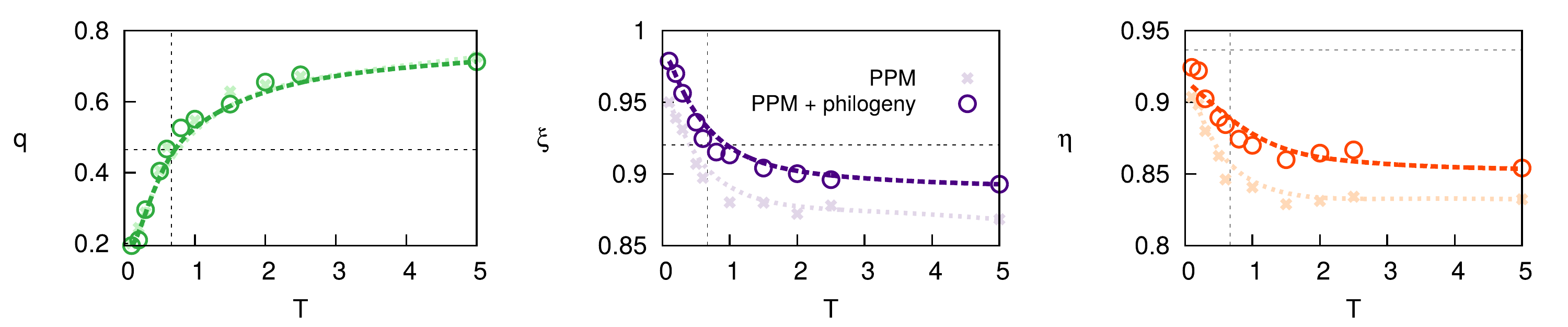}
\end{center}
\caption{Figure analogous to Fig. \ref{fig_q_inter} for the Chesapeake
  Bay food web,\cite{chesapeake1,chesapeake2} but employing the
  modified --rather than the original-- version of the PPM, which
  includes a preference to choose prey which already share some
  predators (or prey).  Results for the original PPM are also reported
  (light colors) for the sake of comparison.  Dotted horizontal lines
  indicate the empirical values as measured in the real food web,
  while vertical ones stand for the value of $T$ providing the best
  fit to $q$. Observe that, while trophic coherence is not
  significantly changed by this modification to the model, the levels
  of intervality ($\xi$ and $\eta$) are clearly augmented by the
  effect of a phylogenetic signal and become closer to the real
  values.  }
\label{fig_philoPPM}
\end{figure*}
Figure \ref{fig_philoPPM} illustrates the performance of this modified
PPM model compared to its original version for the Chesapeake bay food
web. In particular, we generate networks with the modified model
fixing $N$, $L$ and $B$ as in the empirical network, and choose the
optimal value of $T$ (vertical dashed line) that leads to a value of
$q$ as close as possible to the empirical one (horizontal dotted line
in Fig. \ref{fig_philoPPM}).  For this same value of $T$, the obtained
intervality $\xi$ fits quite well with its empirical counterpart
(z-score $ -0.22$).

Observe, that while the dependence of the incoherence parameter, $q$,
on the only parameter, $T$, remains as in the original version, the
intervality $\xi$ obtained with the modified model is higher and much
closer to the empirical value. Similar results are obtained, in
general, for other trophic networks (e.g. for Narragan bay the z-score
is $-0.43$, for St. Marks $-1.06$, and for St. Martin $-0.89$), while
a few exceptions exist (e.g. for the Everglades marshes we obtain
$-6.52$, indicating that in this case our model does not account for the observed intervality). 
Worse agreement is obtained for predator intervality
$\eta$; in the example of Figure \ref{fig_philoPPM}, the corresponding
z-score is $-2.04$. The reason for this is that intervality is
strongly influenced by the degree sequences (see the definitions of
$\xi$ and $\eta$) and the PPM model --as well as its modified
variant-- fit the empirical in-degree distribution, but not the
out-degree one, which is the one affecting $\eta$ the most.

The justification for the proposed combination of trophic
specialization and phylogenetic relations has a clear meaning only in
the case of food webs. We leave as an open question to determine which
mechanisms akin to these ones are at work in systems other than food
webs, or whether there are different mechanisms which have a similar
effect on intervality.

\section{Discussion}
\noindent
The niche concept has been central to ecology for many
decades.\cite{Hutchinson_book,Colwell} McInerny \& Etienne, however,
have recently called into question its prominent role, arguing that
the ecological niche is often not well defined, can create needless
confusion, and may not even be necessary for
ecology.\cite{ditch_niche,pitch_niche} The results we describe here
are relevant mainly for the niche idea as used in food-web modelling.
This approach has its origins in the intervality of food webs first
reported by Cohen,\cite{Cohen} and became firmly established thanks to
the success of Williams \& Martinez's celebrated niche
model.\cite{Williams_niche_model} All subsequent ``structural''
food-web models --i.e. those which attempt to generate realistic
networks without modelling evolutionary or population dynamics
explicitly-- have been based on the niche
model.\cite{Cattin,Dunne_rev,Stouffer_GNM,allesina_model,nicheforever}
Going on this, attempts have been made to identify the niche axis with
some biological magnitude, such as body size, but no such measure has
ever been found to align completely with the orderings of maximum
intervality.\cite{Stouffer_body-size,Zook} It is worth noting that
Rossberg {\it et al.} \cite{Rossberg} were able to generate interval
networks in an $N$-dimensional niche space provided phylogenetic
relationships were present, thus proving that, contrary to general
belief, a one-dimensional space was not a necessary condition for the
emergence of intervality. In their model species evolution is
represented as a random walker in the $N$-dimensional niche space, and
over time species can go extinct or speciate into new ones. As long as
new species are close to the parent ones in niche space, the resulting
networks are highly interval.

We propose here that the intervality of food webs does not justify
invoking hidden or niche dimensions.  We have shown that food webs exhibit as
much predator-intervality as prey-intervality, which seems
incompatible with the interpretation of niche-based models. Moreover,
we find that many different kinds of biological and artificial
networks are as significantly interval as food webs. This means that
either these various networks were all assembled according to the
stringent constraints of a niche-axis analogue --which appears highly
unlikely-- or significant intervality need not be the hallmark of
hidden dimensions.  The latter explanation is further supported by the
strong correlation between intervality and the average proportion of
shared neighbours (complementarity) found across all
networks. Finally, a one-parameter, entirely niche-free network model,
based on the ``preferential preying''\cite{coherence_PNAS} and the
``nested hierarchy''\cite{Cattin} models, which emulates trophic
specialization and phylogenetic relations, yields similar levels of
intervality as those observed in food webs and other networks.

There are two messages to be drawn from this paper. For ecologists, we
cannot say whether our results should change the interpretation of the
niche concept more broadly, but as regards food-web modelling we
believe the case is clear for a new approach. The niche axis --like
Ptolemy's epicycles, phlogiston, caloric, luminiferous aether, and
other constructs eventually found lacking in empirical support--
should be abandoned. This does not mean that Williams \& Martinez's
idea of food-web complexity arising from simple rules is wrong, only
that we must update our concepts of these rules.

For the study of complex networks more generally, these results show
that there are still open questions about the basic mechanisms behind
their formation. Shared neighbours explain $70\%$ of the variance in
intervality, but whence the rest? Do other properties induce such
apparent order, or are there, after all, hidden hierarchies underlying
the architectures of certain complex systems?\cite{Sole,loopines}  The high significance
of topological properties such as intervality and trophic coherence
found in many different kinds of system, from neural networks to
trading relations, and word graphs to gene regulatory networks,
suggests that we, like Darwin admiring the complexity of his entangled
bank, have yet much to learn from ecosystems.

\section{methods} 
\noindent
{ \footnotesize {\textbf{Measuring intervality.}}
  Given a food web and a specified ordering of species $O$, we define
  the prey-intervality, $\xi^O_i$, of each predator species $i$ as the
  largest number of its prey to lie on an unbroken interval divided by the total number of prey it has. The
  prey-intervality of the ordering is then the average over all
  predator species of this value, $\xi^O=\frac{1}{N-B} \sum_i
  \xi^O_i$. The prey-intervality of the food web itself is defined as
  that of the ordering with the highest prey-intervality:
  $\xi=\max\lbrace\xi^O\rbrace_O$. Given that the search cannot be
  made exhaustively, we implement a standard simulated annealing
  algorithm. For this, one begins with a random ordering $O_1$ and
  measures its prey-intervality $\xi^{O_1}$; then two species are
  randomly chosen and their positions in the ordering exchanged,
  yielding a new ordering $O_2$.  The new prey-intervality
  $\xi^{O_2}$ is calculated; and the permutation is 
  accepted with probability
$$
P_{1\rightarrow 2}=\min\left\lbrace \exp\left(\frac{\xi^{O_2}-\xi^{O_1} }{p}\right), 1 \right\rbrace,
$$
where $p$ is a temperature-like parameter, while with probability
1-$P_{1\rightarrow 2}$ the permutation is rejected and the algorithm
is iterated. As is standard, we gradually lower $p$ (starting with
$p=0.01$ and then decreasing by a factor $0.99$ after every $1000$
proposed permutations),
and perform several runs from different initial conditions for each measurement.
In all
cases, the algorithm converges to a unique value of $\xi$ for a given
network. The predator-intervality, $\eta$, is obtained in an analogous
way.  Observe that this measure is slightly different from the one
defined by Stouffer {\it et al.},\cite{Stouffer_robust} which 
is based on the ``generalized niche model'';
but both measures are
strongly correlated (we have found a Pearson's coefficient $r=0.82$
between the two measurements over our network dataset).

{\textbf{Measuring complementarity.}} Given a food web we define its
matrix of shared prey as $M=AA^{T}$, where $A$ is the adjacency
matrix. The element $m_{ij}$ is the number of prey that are shared by
species $i$ and $j$.  Similarly, the matrix of shared predators is
$\hat{M}=A^TA$. The in-complementarity between $i$ and $j$
is defined as $w_{ij}=m_{ij}/max(k_i^{in},k_j^{in})$, and similarly,
for the out-complementarity
$\hat{w}_{ij}=\hat{m}_{ij}/max(k_i^{out},k_j^{out})$.  By averaging
over all possible pairs with at least one common shared in/out
neighbour, we determine the overall complementarities $c_{in}$ and
$c_{out}$, respectively.

{\bf Modified Preferential preying model with phylogenetic constraints.}
\label{appendix_philoppm}
The original preferential preying model was designed to generate
networks with tunable trophic coherence. Here we introduce an
additional mechanism implementing the idea that a predator is likely to choose
prey which are similar, i.e. that share other predators or
other prey.  In the standard PPM one starts with $B$ basal nodes and
adds progressively new species, up to a total of $S$. Each one selects
its prey from existing network nodes, following these rules: i) The
in-degree ($k_{in}$) of species $i$ is selected from a Beta
distribution, so as to obtain on average the same number of links $L$
as the empirical network to be modelled.  ii) The first prey species ($l$) is
randomly selected from the available ones.  iii) Subsequent prey ($j$)
are chosen with a probability $P_{il}$ that decays with the trophic
distance between $j$ and $l$: $$P_{il}\sim \exp(-
\frac{|s_j-s_l|}{T}),$$ where $T$ is a ``temperature'' parameter that
sets the degree of trophic specialization.  In this version of
the PPM, in order to include an additional preference for phylogenetically
related species as done by Cattin {\it et al.}\cite{Cattin}, we modify rule iii) as follows. First
we chose a prey $j$ as above and consider all other possible prey with
the same trophic level (plus/minus $1\%$) and from this group we
randomly select one among the subgroup that already has some predator
with any of the existing prey of $i$. If there is no species obeying
these constraints, then $j$ is selected.

\begin{acknowledgments}
  {\bf Acknowledgments --}
 We acknowledge the Spanish-MINECO grant FIS2013-43201-P (FEDER
  funds) for financial support.
\end{acknowledgments}


\begin{thebibliography}{80}%
\makeatletter
\providecommand \@ifxundefined [1]{%
 \@ifx{#1\undefined}
}%
\providecommand \@ifnum [1]{%
 \ifnum #1\expandafter \@firstoftwo
 \else \expandafter \@secondoftwo
 \fi
}%
\providecommand \@ifx [1]{%
 \ifx #1\expandafter \@firstoftwo
 \else \expandafter \@secondoftwo
 \fi
}%
\providecommand \natexlab [1]{#1}%
\providecommand \enquote  [1]{``#1''}%
\providecommand \bibnamefont  [1]{#1}%
\providecommand \bibfnamefont [1]{#1}%
\providecommand \citenamefont [1]{#1}%
\providecommand \href@noop [0]{\@secondoftwo}%
\providecommand \href [0]{\begingroup \@sanitize@url \@href}%
\providecommand \@href[1]{\@@startlink{#1}\@@href}%
\providecommand \@@href[1]{\endgroup#1\@@endlink}%
\providecommand \@sanitize@url [0]{\catcode `\\12\catcode `\$12\catcode
  `\&12\catcode `\#12\catcode `\^12\catcode `\_12\catcode `\%12\relax}%
\providecommand \@@startlink[1]{}%
\providecommand \@@endlink[0]{}%
\providecommand \url  [0]{\begingroup\@sanitize@url \@url }%
\providecommand \@url [1]{\endgroup\@href {#1}{\urlprefix }}%
\providecommand \urlprefix  [0]{URL }%
\providecommand \Eprint [0]{\href }%
\providecommand \doibase [0]{http://dx.doi.org/}%
\providecommand \selectlanguage [0]{\@gobble}%
\providecommand \bibinfo  [0]{\@secondoftwo}%
\providecommand \bibfield  [0]{\@secondoftwo}%
\providecommand \translation [1]{[#1]}%
\providecommand \BibitemOpen [0]{}%
\providecommand \bibitemStop [0]{}%
\providecommand \bibitemNoStop [0]{.\EOS\space}%
\providecommand \EOS [0]{\spacefactor3000\relax}%
\providecommand \BibitemShut  [1]{\csname bibitem#1\endcsname}%
\let\auto@bib@innerbib\@empty
\bibitem [{\citenamefont {Darwin}(1859)}]{Darwin}%
  \BibitemOpen
  \bibfield  {author} {\bibinfo {author} {\bibfnamefont {C.}~\bibnamefont
  {Darwin}},\ }\href@noop {} {\emph {\bibinfo {title} {On the Origin of
  Species}}}\ (\bibinfo  {publisher} {John Murray},\ \bibinfo {address}
  {London,UK},\ \bibinfo {year} {1859})\BibitemShut {NoStop}%
\bibitem [{\citenamefont {Elton}(1927)}]{Elton}%
  \BibitemOpen
  \bibfield  {author} {\bibinfo {author} {\bibfnamefont {C.~S.}\ \bibnamefont
  {Elton}},\ }\href@noop {} {\emph {\bibinfo {title} {Animal Ecology}}}\
  (\bibinfo  {publisher} {Sidgwick and Jackson},\ \bibinfo {address} {London},\
  \bibinfo {year} {1927})\BibitemShut {NoStop}%
\bibitem [{\citenamefont {Cohen}\ and\ \citenamefont {Newman}(1985)}]{Cohen_1}%
  \BibitemOpen
  \bibfield  {author} {\bibinfo {author} {\bibfnamefont {J.~E.}\ \bibnamefont
  {Cohen}}\ and\ \bibinfo {author} {\bibfnamefont {C.~M.}\ \bibnamefont
  {Newman}},\ }\bibfield  {title} {\enquote {\bibinfo {title} {A stochastic
  theory of community food webs {I}. models and aggregated data},}\ }\href@noop
  {} {\bibfield  {journal} {\bibinfo  {journal} {Proc. R. Soc. London Ser. B.}\
  }\textbf {\bibinfo {volume} {224}},\ \bibinfo {pages} {421--448} (\bibinfo
  {year} {1985})}\BibitemShut {NoStop}%
\bibitem [{\citenamefont {Cohen}(1978)}]{Cohen_book}%
  \BibitemOpen
  \bibfield  {author} {\bibinfo {author} {\bibfnamefont {J.~E.}\ \bibnamefont
  {Cohen}},\ }\href@noop {} {\emph {\bibinfo {title} {Food Webs and Niche
  Space}}}\ (\bibinfo  {publisher} {Princeton Univ. Press},\ \bibinfo {address}
  {Princeton, New Jersey},\ \bibinfo {year} {1978})\BibitemShut {NoStop}%
\bibitem [{\citenamefont {Pimm}(1991)}]{Pimm}%
  \BibitemOpen
  \bibfield  {author} {\bibinfo {author} {\bibfnamefont {S.~L.}\ \bibnamefont
  {Pimm}},\ }\href@noop {} {\emph {\bibinfo {title} {The Balance of Nature?
  Ecological Issues in the Conservation of Species and Communities}}}\
  (\bibinfo  {publisher} {The University of Chicago Press},\ \bibinfo {address}
  {Chicago},\ \bibinfo {year} {1991})\BibitemShut {NoStop}%
\bibitem [{\citenamefont {Drossel}\ and\ \citenamefont
  {McKane}(2003)}]{Drossel}%
  \BibitemOpen
  \bibfield  {author} {\bibinfo {author} {\bibfnamefont {B.}~\bibnamefont
  {Drossel}}\ and\ \bibinfo {author} {\bibfnamefont {A.~J.}\ \bibnamefont
  {McKane}},\ }\href@noop {} {\emph {\bibinfo {title} {“Modelling Food
  Webs”, in {A} {H}andbook of {G}raphs and {N}etworks: {F}rom the {G}enome to
  the {I}nternet}}}\ (\bibinfo  {publisher} {Wiley-VCH},\ \bibinfo {address}
  {Berlin},\ \bibinfo {year} {2003})\BibitemShut {NoStop}%
\bibitem [{\citenamefont {Dunne}(2006)}]{Dunne_rev}%
  \BibitemOpen
  \bibfield  {author} {\bibinfo {author} {\bibfnamefont {J.~A.}\ \bibnamefont
  {Dunne}},\ }\href@noop {} {\emph {\bibinfo {title} {The network structure of
  food webs, in {E}cological {N}etworks: {L}inking Structure to Dynamics in
  Food Webs}}},\ edited by\ \bibinfo {editor} {\bibfnamefont {M.}~\bibnamefont
  {Pascual}}\ and\ \bibinfo {editor} {\bibfnamefont {e.}~\bibnamefont
  {J.A.~Dunne}}\ (\bibinfo  {publisher} {Oxford University Press},\ \bibinfo
  {address} {Oxford, UK},\ \bibinfo {year} {2006})\BibitemShut {NoStop}%
\bibitem [{\citenamefont {Sol\'e}\ and\ \citenamefont
  {Bascompte}(2006)}]{Jordi_book}%
  \BibitemOpen
  \bibfield  {author} {\bibinfo {author} {\bibfnamefont {R.~V.}\ \bibnamefont
  {Sol\'e}}\ and\ \bibinfo {author} {\bibfnamefont {J.}~\bibnamefont
  {Bascompte}},\ }\href@noop {} {\emph {\bibinfo {title} {Self-Organization in
  Complex Ecosystems}}}\ (\bibinfo  {publisher} {Princeton University Press},\
  \bibinfo {address} {Princeton, USA},\ \bibinfo {year} {2006})\BibitemShut
  {NoStop}%
\bibitem [{\citenamefont {Camacho}, \citenamefont {Guimer\'a},\ and\
  \citenamefont {Amaral}(2002)}]{guimera_foodweb}%
  \BibitemOpen
  \bibfield  {author} {\bibinfo {author} {\bibfnamefont {J.}~\bibnamefont
  {Camacho}}, \bibinfo {author} {\bibfnamefont {R.}~\bibnamefont {Guimer\'a}},
  \ and\ \bibinfo {author} {\bibfnamefont {L.~A.~N.}\ \bibnamefont {Amaral}},\
  }\bibfield  {title} {\enquote {\bibinfo {title} {Robust patterns in food web
  structure},}\ }\href@noop {} {\bibfield  {journal} {\bibinfo  {journal}
  {Phys. Rev. Lett.}\ }\textbf {\bibinfo {volume} {88}},\ \bibinfo {pages}
  {228102} (\bibinfo {year} {2002})}\BibitemShut {NoStop}%
\bibitem [{\citenamefont {Sol\'e}\ and\ \citenamefont
  {Montoya}(2001)}]{sole_2001}%
  \BibitemOpen
  \bibfield  {author} {\bibinfo {author} {\bibfnamefont {R.~V.}\ \bibnamefont
  {Sol\'e}}\ and\ \bibinfo {author} {\bibfnamefont {M.}~\bibnamefont
  {Montoya}},\ }\bibfield  {title} {\enquote {\bibinfo {title} {Complexity and
  fragility in ecological networks},}\ }\href@noop {} {\bibfield  {journal}
  {\bibinfo  {journal} {Proc. R. Soc. Lond. B}\ }\textbf {\bibinfo {volume}
  {268}},\ \bibinfo {pages} {2039–204} (\bibinfo {year} {2001})}\BibitemShut
  {NoStop}%
\bibitem [{\citenamefont {De~Vos}\ \emph {et~al.}(2015)\citenamefont {De~Vos},
  \citenamefont {Joppa}, \citenamefont {Gittleman}, \citenamefont {Stephens},\
  and\ \citenamefont {Pimm}}]{de2015estimating}%
  \BibitemOpen
  \bibfield  {author} {\bibinfo {author} {\bibfnamefont {J.~M.}\ \bibnamefont
  {De~Vos}}, \bibinfo {author} {\bibfnamefont {L.~N.}\ \bibnamefont {Joppa}},
  \bibinfo {author} {\bibfnamefont {J.~L.}\ \bibnamefont {Gittleman}}, \bibinfo
  {author} {\bibfnamefont {P.~R.}\ \bibnamefont {Stephens}}, \ and\ \bibinfo
  {author} {\bibfnamefont {S.~L.}\ \bibnamefont {Pimm}},\ }\bibfield  {title}
  {\enquote {\bibinfo {title} {Estimating the normal background rate of species
  extinction},}\ }\href@noop {} {\bibfield  {journal} {\bibinfo  {journal}
  {Conservation Biology}\ }\textbf {\bibinfo {volume} {29}},\ \bibinfo {pages}
  {452--462} (\bibinfo {year} {2015})}\BibitemShut {NoStop}%
\bibitem [{\citenamefont {Cohen}(1977)}]{Cohen}%
  \BibitemOpen
  \bibfield  {author} {\bibinfo {author} {\bibfnamefont {J.~E.}\ \bibnamefont
  {Cohen}},\ }\bibfield  {title} {\enquote {\bibinfo {title} {Food webs and the
  dimensionality of trophic niche space},}\ }\href@noop {} {\bibfield
  {journal} {\bibinfo  {journal} {Proc. Natl. Acad. Sci. USA}\ }\textbf
  {\bibinfo {volume} {74}},\ \bibinfo {pages} {4533--4563} (\bibinfo {year}
  {1977})}\BibitemShut {NoStop}%
\bibitem [{\citenamefont {Grinnell}(1917)}]{Grinnell}%
  \BibitemOpen
  \bibfield  {author} {\bibinfo {author} {\bibfnamefont {J.}~\bibnamefont
  {Grinnell}},\ }\bibfield  {title} {\enquote {\bibinfo {title} {The
  niche-relationships of the {C}alifornia {T}hrasher},}\ }\href@noop {}
  {\bibfield  {journal} {\bibinfo  {journal} {Auk}\ }\textbf {\bibinfo {volume}
  {34}},\ \bibinfo {pages} {427--433} (\bibinfo {year} {1917})}\BibitemShut
  {NoStop}%
\bibitem [{\citenamefont {Hutchinson}(1957)}]{Hutchinson}%
  \BibitemOpen
  \bibfield  {author} {\bibinfo {author} {\bibfnamefont {G.~E.}\ \bibnamefont
  {Hutchinson}},\ }\bibfield  {title} {\enquote {\bibinfo {title} {Concluding
  remarks},}\ }\href@noop {} {\bibfield  {journal} {\bibinfo  {journal} {Cold
  Springs Harbor Symp. Quant. Biol.}\ }\textbf {\bibinfo {volume} {22}},\
  \bibinfo {pages} {415--427} (\bibinfo {year} {1957})}\BibitemShut {NoStop}%
\bibitem [{\citenamefont {Williams}\ and\ \citenamefont
  {Martinez}(2000)}]{Williams_niche_model}%
  \BibitemOpen
  \bibfield  {author} {\bibinfo {author} {\bibfnamefont {R.~J.}\ \bibnamefont
  {Williams}}\ and\ \bibinfo {author} {\bibfnamefont {N.~D.}\ \bibnamefont
  {Martinez}},\ }\bibfield  {title} {\enquote {\bibinfo {title} {Simple rules
  yield complex food webs},}\ }\href@noop {} {\bibfield  {journal} {\bibinfo
  {journal} {Nature}\ }\textbf {\bibinfo {volume} {404}},\ \bibinfo {pages}
  {180--183} (\bibinfo {year} {2000})}\BibitemShut {NoStop}%
\bibitem [{\citenamefont {Williams}\ and\ \citenamefont
  {Martinez}(2008)}]{nicheforever}%
  \BibitemOpen
  \bibfield  {author} {\bibinfo {author} {\bibfnamefont {R.~J.}\ \bibnamefont
  {Williams}}\ and\ \bibinfo {author} {\bibfnamefont {N.~D.}\ \bibnamefont
  {Martinez}},\ }\bibfield  {title} {\enquote {\bibinfo {title} {Success and
  its limits among structural models of complex food webs},}\ }\href@noop {}
  {\bibfield  {journal} {\bibinfo  {journal} {Journal of Animal Ecology}\
  }\textbf {\bibinfo {volume} {77}},\ \bibinfo {pages} {512--519} (\bibinfo
  {year} {2008})}\BibitemShut {NoStop}%
\bibitem [{\citenamefont {Guill}\ and\ \citenamefont
  {Drossel}(2008)}]{drosselniche}%
  \BibitemOpen
  \bibfield  {author} {\bibinfo {author} {\bibfnamefont {C.}~\bibnamefont
  {Guill}}\ and\ \bibinfo {author} {\bibfnamefont {B.}~\bibnamefont
  {Drossel}},\ }\bibfield  {title} {\enquote {\bibinfo {title} {Emergence of
  complexity in evolving niche-model food webs},}\ }\href@noop {} {\bibfield
  {journal} {\bibinfo  {journal} {Journal of Theoretical Biology}\ }\textbf
  {\bibinfo {volume} {251}},\ \bibinfo {pages} {108–120} (\bibinfo {year}
  {2008})}\BibitemShut {NoStop}%
\bibitem [{\citenamefont {Stouffer}, \citenamefont {Camacho},\ and\
  \citenamefont {Amaral}(2006)}]{Stouffer_robust}%
  \BibitemOpen
  \bibfield  {author} {\bibinfo {author} {\bibfnamefont {D.~B.}\ \bibnamefont
  {Stouffer}}, \bibinfo {author} {\bibfnamefont {J.}~\bibnamefont {Camacho}}, \
  and\ \bibinfo {author} {\bibfnamefont {L.~A.~N.}\ \bibnamefont {Amaral}},\
  }\bibfield  {title} {\enquote {\bibinfo {title} {A robust measure of food web
  intervality},}\ }\href@noop {} {\bibfield  {journal} {\bibinfo  {journal}
  {Proc. Natl. Acad. Sci. USA}\ }\textbf {\bibinfo {volume} {103}},\ \bibinfo
  {pages} {19015--19020} (\bibinfo {year} {2006})}\BibitemShut {NoStop}%
\bibitem [{\citenamefont {Zook}\ \emph {et~al.}(2011)\citenamefont {Zook},
  \citenamefont {Ekl\"of}, \citenamefont {Jacob},\ and\ \citenamefont
  {Allesina}}]{Zook}%
  \BibitemOpen
  \bibfield  {author} {\bibinfo {author} {\bibfnamefont {A.~E.}\ \bibnamefont
  {Zook}}, \bibinfo {author} {\bibfnamefont {A.}~\bibnamefont {Ekl\"of}},
  \bibinfo {author} {\bibfnamefont {U.}~\bibnamefont {Jacob}}, \ and\ \bibinfo
  {author} {\bibfnamefont {S.}~\bibnamefont {Allesina}},\ }\bibfield  {title}
  {\enquote {\bibinfo {title} {Food webs: {O}rdering species according to body
  size yields high degree of intervality},}\ }\href@noop {} {\bibfield
  {journal} {\bibinfo  {journal} {Journal of Theoretical Biology}\ }\textbf
  {\bibinfo {volume} {271}},\ \bibinfo {pages} {106--113} (\bibinfo {year}
  {2011})}\BibitemShut {NoStop}%
\bibitem [{\citenamefont {Patricio}(2000)}]{mondego}%
  \BibitemOpen
  \bibfield  {author} {\bibinfo {author} {\bibfnamefont {J.}~\bibnamefont
  {Patricio}},\ }\bibfield  {title} {\enquote {\bibinfo {title} {Network
  analysis of trophic dynamics in south florida ecosystems, fy 99: The
  graminoid ecosystem.}}\ }\href@noop {} {\bibfield  {journal} {\bibinfo
  {journal} {Master's Thesis. University of Coimbra, Coimbra, Portugal}\ }
  (\bibinfo {year} {2000})}\BibitemShut {NoStop}%
\bibitem [{\citenamefont {Cattin}\ \emph {et~al.}(2004)\citenamefont {Cattin},
  \citenamefont {Bersier}, \citenamefont {Banasek-Richter}, \citenamefont
  {Baltensperger},\ and\ \citenamefont {Gabriel}}]{Cattin}%
  \BibitemOpen
  \bibfield  {author} {\bibinfo {author} {\bibfnamefont {M.~F.}\ \bibnamefont
  {Cattin}}, \bibinfo {author} {\bibfnamefont {L.~F.}\ \bibnamefont {Bersier}},
  \bibinfo {author} {\bibfnamefont {C.}~\bibnamefont {Banasek-Richter}},
  \bibinfo {author} {\bibfnamefont {R.}~\bibnamefont {Baltensperger}}, \ and\
  \bibinfo {author} {\bibfnamefont {J.~P.}\ \bibnamefont {Gabriel}},\
  }\bibfield  {title} {\enquote {\bibinfo {title} {Phylogenetic constraints and
  adaptation explain food-web structure},}\ }\href@noop {} {\bibfield
  {journal} {\bibinfo  {journal} {Nature}\ }\textbf {\bibinfo {volume} {427}},\
  \bibinfo {pages} {835--9} (\bibinfo {year} {2004})}\BibitemShut {NoStop}%
\bibitem [{\citenamefont {Girvan}\ and\ \citenamefont
  {Newman}(2002)}]{Newman_compartments}%
  \BibitemOpen
  \bibfield  {author} {\bibinfo {author} {\bibfnamefont {M.}~\bibnamefont
  {Girvan}}\ and\ \bibinfo {author} {\bibfnamefont {M.~E.~J.}\ \bibnamefont
  {Newman}},\ }\bibfield  {title} {\enquote {\bibinfo {title} {Community
  structure in social and biological networks},}\ }\href@noop {} {\bibfield
  {journal} {\bibinfo  {journal} {Proc. Natl. Acad. Sci. USA}\ }\textbf
  {\bibinfo {volume} {99}},\ \bibinfo {pages} {7821--7826} (\bibinfo {year}
  {2002})}\BibitemShut {NoStop}%
\bibitem [{\citenamefont {Johnson}, \citenamefont {Dom\'inguez-Garc\'ia},\ and\
  \citenamefont {Mu{\~n}oz}(2013)}]{plos_nestedness}%
  \BibitemOpen
  \bibfield  {author} {\bibinfo {author} {\bibfnamefont {S.}~\bibnamefont
  {Johnson}}, \bibinfo {author} {\bibfnamefont {V.}~\bibnamefont
  {Dom\'inguez-Garc\'ia}}, \ and\ \bibinfo {author} {\bibfnamefont {M.~A.}\
  \bibnamefont {Mu{\~n}oz}},\ }\bibfield  {title} {\enquote {\bibinfo {title}
  {Factors determining nestedness in complex networks},}\ }\href@noop {}
  {\bibfield  {journal} {\bibinfo  {journal} {PloS ONE}\ }\textbf {\bibinfo
  {volume} {8}},\ \bibinfo {pages} {e74025} (\bibinfo {year}
  {2013})}\BibitemShut {NoStop}%
\bibitem [{\citenamefont {Stouffer}\ \emph {et~al.}(2005)\citenamefont
  {Stouffer}, \citenamefont {Camacho}, \citenamefont {Guimer\`a}, \citenamefont
  {Ng},\ and\ \citenamefont {Amaral}}]{Stouffer_GNM}%
  \BibitemOpen
  \bibfield  {author} {\bibinfo {author} {\bibfnamefont {D.~B.}\ \bibnamefont
  {Stouffer}}, \bibinfo {author} {\bibfnamefont {J.}~\bibnamefont {Camacho}},
  \bibinfo {author} {\bibfnamefont {R.}~\bibnamefont {Guimer\`a}}, \bibinfo
  {author} {\bibfnamefont {C.~A.}\ \bibnamefont {Ng}}, \ and\ \bibinfo {author}
  {\bibfnamefont {L.~A.~N.}\ \bibnamefont {Amaral}},\ }\bibfield  {title}
  {\enquote {\bibinfo {title} {Quantitative patterns in the structure of model
  and empirical food webs},}\ }\href@noop {} {\bibfield  {journal} {\bibinfo
  {journal} {Ecology}\ }\textbf {\bibinfo {volume} {86}},\ \bibinfo {pages}
  {1301–1311} (\bibinfo {year} {2005})}\BibitemShut {NoStop}%
\bibitem [{\citenamefont {Allesina}, \citenamefont {Alonso},\ and\
  \citenamefont {Pascual}(2008)}]{allesina_model}%
  \BibitemOpen
  \bibfield  {author} {\bibinfo {author} {\bibfnamefont {S.}~\bibnamefont
  {Allesina}}, \bibinfo {author} {\bibfnamefont {D.}~\bibnamefont {Alonso}}, \
  and\ \bibinfo {author} {\bibfnamefont {M.}~\bibnamefont {Pascual}},\
  }\bibfield  {title} {\enquote {\bibinfo {title} {A general model for food web
  structure},}\ }\href@noop {} {\bibfield  {journal} {\bibinfo  {journal}
  {Science}\ }\textbf {\bibinfo {volume} {320}},\ \bibinfo {pages} {658--661}
  (\bibinfo {year} {2008})}\BibitemShut {NoStop}%
\bibitem [{\citenamefont {Johnson}\ \emph {et~al.}(2014)\citenamefont
  {Johnson}, \citenamefont {Dom\'inguez-Garc\'ia}, \citenamefont {Donetti},\
  and\ \citenamefont {Mu\~noz}}]{coherence_PNAS}%
  \BibitemOpen
  \bibfield  {author} {\bibinfo {author} {\bibfnamefont {S.}~\bibnamefont
  {Johnson}}, \bibinfo {author} {\bibfnamefont {V.}~\bibnamefont
  {Dom\'inguez-Garc\'ia}}, \bibinfo {author} {\bibfnamefont {L.}~\bibnamefont
  {Donetti}}, \ and\ \bibinfo {author} {\bibfnamefont {M.~A.}\ \bibnamefont
  {Mu\~noz}},\ }\bibfield  {title} {\enquote {\bibinfo {title} {Trophic
  coherence determines food-web stability},}\ }\href {\doibase
  10.1073/pnas.1409077111} {\bibfield  {journal} {\bibinfo  {journal} {Proc.
  Natl. Acad. Sci. USA}\ }\textbf {\bibinfo {volume} {111}},\ \bibinfo {pages}
  {17923--17928} (\bibinfo {year} {2014})}\BibitemShut {NoStop}%
\bibitem [{\citenamefont {Levine}(1980)}]{Levine_levels}%
  \BibitemOpen
  \bibfield  {author} {\bibinfo {author} {\bibfnamefont {S.}~\bibnamefont
  {Levine}},\ }\bibfield  {title} {\enquote {\bibinfo {title} {Several measures
  of trophic structure applicable to complex food webs},}\ }\href@noop {}
  {\bibfield  {journal} {\bibinfo  {journal} {J. Theor. Biol.}\ }\textbf
  {\bibinfo {volume} {83}},\ \bibinfo {pages} {195--207} (\bibinfo {year}
  {1980})}\BibitemShut {NoStop}%
\bibitem [{\citenamefont {Dom\'inguez-Garc\'ia}, \citenamefont {Pigolotti},\
  and\ \citenamefont {Mu{\~n}oz}(2014)}]{loopines}%
  \BibitemOpen
  \bibfield  {author} {\bibinfo {author} {\bibfnamefont {V.}~\bibnamefont
  {Dom\'inguez-Garc\'ia}}, \bibinfo {author} {\bibfnamefont {S.}~\bibnamefont
  {Pigolotti}}, \ and\ \bibinfo {author} {\bibfnamefont {M.~A.}\ \bibnamefont
  {Mu{\~n}oz}},\ }\bibfield  {title} {\enquote {\bibinfo {title} {Inherent
  directionality explains the lack of feedback loops in empirical networks},}\
  }\href@noop {} {\bibfield  {journal} {\bibinfo  {journal} {Scientific
  Reports}\ }\textbf {\bibinfo {volume} {4}},\ \bibinfo {pages} {7497}
  (\bibinfo {year} {2014})}\BibitemShut {NoStop}%
\bibitem [{\citenamefont {Johnson}\ and\ \citenamefont
  {Jones}(2015)}]{Johnson_spectra}%
  \BibitemOpen
  \bibfield  {author} {\bibinfo {author} {\bibfnamefont {S.}~\bibnamefont
  {Johnson}}\ and\ \bibinfo {author} {\bibfnamefont {N.~S.}\ \bibnamefont
  {Jones}},\ }\bibfield  {title} {\enquote {\bibinfo {title} {Spectra and cycle
  structure of trophically coherent graphs},}\ }\href@noop {} {\bibfield
  {journal} {\bibinfo  {journal} {arXiv:1505.07332}\ } (\bibinfo {year}
  {2015})}\BibitemShut {NoStop}%
\bibitem [{\citenamefont {Dom{\'\i}nguez-Garc{\'\i}a}\ and\ \citenamefont
  {Mu{\~n}oz}(2015)}]{dominguez2015ranking}%
  \BibitemOpen
  \bibfield  {author} {\bibinfo {author} {\bibfnamefont {V.}~\bibnamefont
  {Dom{\'\i}nguez-Garc{\'\i}a}}\ and\ \bibinfo {author} {\bibfnamefont {M.~A.}\
  \bibnamefont {Mu{\~n}oz}},\ }\bibfield  {title} {\enquote {\bibinfo {title}
  {Ranking species in mutualistic networks},}\ }\href@noop {} {\bibfield
  {journal} {\bibinfo  {journal} {Scientific reports}\ }\textbf {\bibinfo
  {volume} {5}} (\bibinfo {year} {2015})}\BibitemShut {NoStop}%
\bibitem [{\citenamefont {May}(1972)}]{May}%
  \BibitemOpen
  \bibfield  {author} {\bibinfo {author} {\bibfnamefont {R.~M.}\ \bibnamefont
  {May}},\ }\bibfield  {title} {\enquote {\bibinfo {title} {Will a large
  complex system be stable?}}\ }\href@noop {} {\bibfield  {journal} {\bibinfo
  {journal} {Nature}\ }\textbf {\bibinfo {volume} {238}},\ \bibinfo {pages}
  {413} (\bibinfo {year} {1972})}\BibitemShut {NoStop}%
\bibitem [{\citenamefont {McCann}(2000)}]{McCann}%
  \BibitemOpen
  \bibfield  {author} {\bibinfo {author} {\bibfnamefont {K.~S.}\ \bibnamefont
  {McCann}},\ }\bibfield  {title} {\enquote {\bibinfo {title} {The
  diversity-stability debate},}\ }\href@noop {} {\bibfield  {journal} {\bibinfo
   {journal} {Nature}\ }\textbf {\bibinfo {volume} {405}},\ \bibinfo {pages}
  {228--33} (\bibinfo {year} {2000})}\BibitemShut {NoStop}%
\bibitem [{\citenamefont {Klaise}\ and\ \citenamefont
  {Johnson}(2016)}]{Janis_chaos}%
  \BibitemOpen
  \bibfield  {author} {\bibinfo {author} {\bibfnamefont {J.}~\bibnamefont
  {Klaise}}\ and\ \bibinfo {author} {\bibfnamefont {S.}~\bibnamefont
  {Johnson}},\ }\bibfield  {title} {\enquote {\bibinfo {title} {From neurons to
  epidemics: {H}ow trophic coherence affects spreading processes},}\
  }\href@noop {} {\bibfield  {journal} {\bibinfo  {journal} {arXiv:1603.00670}\
  } (\bibinfo {year} {2016})}\BibitemShut {NoStop}%
\bibitem [{Note1()}]{Note1}%
  \BibitemOpen
  \bibinfo {note} {This is the model we use to generate coherent networks in
  this work; however, we note that Klaise \& Johnson\cite {Janis_chaos} propose
  a slightly different version of the PPM, the main difference being that at
  high $T$ their model limits in random graphs instead of acyclic cascade model
  networks.}\BibitemShut {Stop}%
\bibitem [{\citenamefont {Ulanowicz}\ and\ \citenamefont
  {Baird}(1999)}]{chesapeake1}%
  \BibitemOpen
  \bibfield  {author} {\bibinfo {author} {\bibfnamefont {R.~E.}\ \bibnamefont
  {Ulanowicz}}\ and\ \bibinfo {author} {\bibfnamefont {D.}~\bibnamefont
  {Baird}},\ }\bibfield  {title} {\enquote {\bibinfo {title} {Nutrient controls
  on ecosystem dynamics: the chesapeake mesohaline community},}\ }\href@noop {}
  {\bibfield  {journal} {\bibinfo  {journal} {Journal of Marine Systems}\
  }\textbf {\bibinfo {volume} {19}},\ \bibinfo {pages} {159 -- 172} (\bibinfo
  {year} {1999})}\BibitemShut {NoStop}%
\bibitem [{\citenamefont {Abarca-Arenas}\ and\ \citenamefont
  {Ulanowicz}(2002)}]{chesapeake2}%
  \BibitemOpen
  \bibfield  {author} {\bibinfo {author} {\bibfnamefont {L.~G.}\ \bibnamefont
  {Abarca-Arenas}}\ and\ \bibinfo {author} {\bibfnamefont {R.~E.}\ \bibnamefont
  {Ulanowicz}},\ }\bibfield  {title} {\enquote {\bibinfo {title} {The effects
  of taxonomic aggregation on network analysis},}\ }\href@noop {} {\bibfield
  {journal} {\bibinfo  {journal} {Ecological Modelling}\ }\textbf {\bibinfo
  {volume} {149}},\ \bibinfo {pages} {285 -- 296} (\bibinfo {year}
  {2002})}\BibitemShut {NoStop}%
\bibitem [{\citenamefont {Capit\'an}, \citenamefont {Arenas},\ and\
  \citenamefont {Guimer\'a}(2013)}]{arenas_model}%
  \BibitemOpen
  \bibfield  {author} {\bibinfo {author} {\bibfnamefont {J.}~\bibnamefont
  {Capit\'an}}, \bibinfo {author} {\bibfnamefont {A.}~\bibnamefont {Arenas}}, \
  and\ \bibinfo {author} {\bibfnamefont {R.}~\bibnamefont {Guimer\'a}},\
  }\bibfield  {title} {\enquote {\bibinfo {title} {Degree of intervality of
  food webs: From body-size data to models},}\ }\href@noop {} {\bibfield
  {journal} {\bibinfo  {journal} {Journal of Theoretical Biology}\ }\textbf
  {\bibinfo {volume} {334}},\ \bibinfo {pages} {35--44} (\bibinfo {year}
  {2013})}\BibitemShut {NoStop}%
\bibitem [{\citenamefont {Ekl\"of}\ and\ \citenamefont
  {Stouffer}(2015)}]{eklof_philo}%
  \BibitemOpen
  \bibfield  {author} {\bibinfo {author} {\bibfnamefont {A.}~\bibnamefont
  {Ekl\"of}}\ and\ \bibinfo {author} {\bibfnamefont {D.~B.}\ \bibnamefont
  {Stouffer}},\ }\bibfield  {title} {\enquote {\bibinfo {title} {The
  phylogenetic component of food web structure and intervality},}\ }\href@noop
  {} {\bibfield  {journal} {\bibinfo  {journal} {Theor Ecol}\ } (\bibinfo
  {year} {2015})}\BibitemShut {NoStop}%
\bibitem [{\citenamefont {Mouillot}, \citenamefont {Krasnov},\ and\
  \citenamefont {Poulin}(2008)}]{pulin_philo}%
  \BibitemOpen
  \bibfield  {author} {\bibinfo {author} {\bibfnamefont {D.}~\bibnamefont
  {Mouillot}}, \bibinfo {author} {\bibfnamefont {B.}~\bibnamefont {Krasnov}}, \
  and\ \bibinfo {author} {\bibfnamefont {R.}~\bibnamefont {Poulin}},\
  }\bibfield  {title} {\enquote {\bibinfo {title} {High intervality explained
  by phylogenetic constraints in host-parasite webs},}\ }\href@noop {}
  {\bibfield  {journal} {\bibinfo  {journal} {Ecology}\ }\textbf {\bibinfo
  {volume} {89(7)}},\ \bibinfo {pages} {2043–2051} (\bibinfo {year}
  {2008})}\BibitemShut {NoStop}%
\bibitem [{\citenamefont {Hutchinson}(1978)}]{Hutchinson_book}%
  \BibitemOpen
  \bibfield  {author} {\bibinfo {author} {\bibfnamefont {G.~E.}\ \bibnamefont
  {Hutchinson}},\ }\href@noop {} {\emph {\bibinfo {title} {An Introduction to
  Population Biology}}}\ (\bibinfo  {publisher} {Yale Univ Press},\ \bibinfo
  {address} {New Haven, CT, USA.},\ \bibinfo {year} {1978})\BibitemShut
  {NoStop}%
\bibitem [{\citenamefont {Colwell}\ and\ \citenamefont
  {Rangel}(2009)}]{Colwell}%
  \BibitemOpen
  \bibfield  {author} {\bibinfo {author} {\bibfnamefont {R.~K.}\ \bibnamefont
  {Colwell}}\ and\ \bibinfo {author} {\bibfnamefont {T.~F.}\ \bibnamefont
  {Rangel}},\ }\bibfield  {title} {\enquote {\bibinfo {title} {Hutchinson's
  duality: The once and future niche},}\ }\href@noop {} {\bibfield  {journal}
  {\bibinfo  {journal} {Proc. Natl. Acad. Sci. USA}\ }\textbf {\bibinfo
  {volume} {106}},\ \bibinfo {pages} {19651} (\bibinfo {year}
  {2009})}\BibitemShut {NoStop}%
\bibitem [{\citenamefont {McInerny}\ and\ \citenamefont
  {Etienne}(2012{\natexlab{a}})}]{ditch_niche}%
  \BibitemOpen
  \bibfield  {author} {\bibinfo {author} {\bibfnamefont {G.}~\bibnamefont
  {McInerny}}\ and\ \bibinfo {author} {\bibfnamefont {R.}~\bibnamefont
  {Etienne}},\ }\bibfield  {title} {\enquote {\bibinfo {title} {Ditch the niche
  – is the niche a useful concept in ecology or species distribution
  modelling?}}\ }\href@noop {} {\bibfield  {journal} {\bibinfo  {journal} {J.
  Biogeogr.}\ }\textbf {\bibinfo {volume} {39}},\ \bibinfo {pages}
  {2096–2102} (\bibinfo {year} {2012}{\natexlab{a}})}\BibitemShut {NoStop}%
\bibitem [{\citenamefont {McInerny}\ and\ \citenamefont
  {Etienne}(2012{\natexlab{b}})}]{pitch_niche}%
  \BibitemOpen
  \bibfield  {author} {\bibinfo {author} {\bibfnamefont {G.}~\bibnamefont
  {McInerny}}\ and\ \bibinfo {author} {\bibfnamefont {R.}~\bibnamefont
  {Etienne}},\ }\bibfield  {title} {\enquote {\bibinfo {title} {Pitch the niche
  – taking responsibility for the concepts we use in ecology and species
  distribution modelling},}\ }\href@noop {} {\bibfield  {journal} {\bibinfo
  {journal} {J. Biogeogr.}\ }\textbf {\bibinfo {volume} {39}},\ \bibinfo
  {pages} {2112–2118} (\bibinfo {year} {2012}{\natexlab{b}})}\BibitemShut
  {NoStop}%
\bibitem [{\citenamefont {Stouffer}, \citenamefont {Rezende},\ and\
  \citenamefont {Amaral}(2011)}]{Stouffer_body-size}%
  \BibitemOpen
  \bibfield  {author} {\bibinfo {author} {\bibfnamefont {D.~B.}\ \bibnamefont
  {Stouffer}}, \bibinfo {author} {\bibfnamefont {E.~L.}\ \bibnamefont
  {Rezende}}, \ and\ \bibinfo {author} {\bibfnamefont {L.~A.~N.}\ \bibnamefont
  {Amaral}},\ }\bibfield  {title} {\enquote {\bibinfo {title} {The role of body
  mass in diet contiguity and food-web structure},}\ }\href@noop {} {\bibfield
  {journal} {\bibinfo  {journal} {J. Anim. Ecol.}\ }\textbf {\bibinfo {volume}
  {80}},\ \bibinfo {pages} {632--639} (\bibinfo {year} {2011})}\BibitemShut
  {NoStop}%
\bibitem [{\citenamefont {Rossberg}, \citenamefont {Br\"annstr\"om},\ and\
  \citenamefont {Dieckmann}(2010)}]{Rossberg}%
  \BibitemOpen
  \bibfield  {author} {\bibinfo {author} {\bibfnamefont {A.~G.}\ \bibnamefont
  {Rossberg}}, \bibinfo {author} {\bibfnamefont {A.}~\bibnamefont
  {Br\"annstr\"om}}, \ and\ \bibinfo {author} {\bibfnamefont {U.}~\bibnamefont
  {Dieckmann}},\ }\bibfield  {title} {\enquote {\bibinfo {title} {Food-web
  structure in low- and high-dimensional trophic niche spaces},}\ }\href@noop
  {} {\bibfield  {journal} {\bibinfo  {journal} {Journal of The Royal Society
  Interface}\ } (\bibinfo {year} {2010})}\BibitemShut {NoStop}%
\bibitem [{\citenamefont {Corominas-Murtra}\ \emph {et~al.}(2013)\citenamefont
  {Corominas-Murtra}, \citenamefont {Go{\~n}i}, \citenamefont {Sol{\'e}},\ and\
  \citenamefont {Rodríguez-Caso}}]{Sole}%
  \BibitemOpen
  \bibfield  {author} {\bibinfo {author} {\bibfnamefont {B.}~\bibnamefont
  {Corominas-Murtra}}, \bibinfo {author} {\bibfnamefont {J.}~\bibnamefont
  {Go{\~n}i}}, \bibinfo {author} {\bibfnamefont {R.~V.}\ \bibnamefont
  {Sol{\'e}}}, \ and\ \bibinfo {author} {\bibfnamefont {C.}~\bibnamefont
  {Rodríguez-Caso}},\ }\bibfield  {title} {\enquote {\bibinfo {title} {On the
  origins of hierarchy in complex networks},}\ }\href@noop {} {\bibfield
  {journal} {\bibinfo  {journal} {Proc. Natl. Acad. Sci. USA}\ }\textbf
  {\bibinfo {volume} {103}},\ \bibinfo {pages} {13316–13321} (\bibinfo {year}
  {2013})}\BibitemShut {NoStop}%
\bibitem [{\citenamefont {Jaarsma}\ \emph {et~al.}(1998)\citenamefont
  {Jaarsma}, \citenamefont {de~Boer}, \citenamefont {Townsend}, \citenamefont
  {Thompson},\ and\ \citenamefont {Edwards}}]{streams5}%
  \BibitemOpen
  \bibfield  {author} {\bibinfo {author} {\bibfnamefont {N.~G.}\ \bibnamefont
  {Jaarsma}}, \bibinfo {author} {\bibfnamefont {S.~M.}\ \bibnamefont
  {de~Boer}}, \bibinfo {author} {\bibfnamefont {C.~R.}\ \bibnamefont
  {Townsend}}, \bibinfo {author} {\bibfnamefont {R.~M.}\ \bibnamefont
  {Thompson}}, \ and\ \bibinfo {author} {\bibfnamefont {E.~D.}\ \bibnamefont
  {Edwards}},\ }\bibfield  {title} {\enquote {\bibinfo {title} {Characterising
  food‐webs in two new zealand streams},}\ }\href@noop {} {\bibfield
  {journal} {\bibinfo  {journal} {New Zealand Journal of Marine and Freshwater
  Research}\ }\textbf {\bibinfo {volume} {32}},\ \bibinfo {pages} {271--286}
  (\bibinfo {year} {1998})}\BibitemShut {NoStop}%
\bibitem [{\citenamefont {Thompson}\ and\ \citenamefont
  {Townsend}(2005)}]{streams6}%
  \BibitemOpen
  \bibfield  {author} {\bibinfo {author} {\bibfnamefont {R.~M.}\ \bibnamefont
  {Thompson}}\ and\ \bibinfo {author} {\bibfnamefont {C.~R.}\ \bibnamefont
  {Townsend}},\ }\bibfield  {title} {\enquote {\bibinfo {title} {Energy
  availability, spatial heterogeneity and ecosystem size predict food-web
  structure in stream},}\ }\href@noop {} {\bibfield  {journal} {\bibinfo
  {journal} {Oikos}\ }\textbf {\bibinfo {volume} {108}},\ \bibinfo {pages}
  {137–148} (\bibinfo {year} {2005})}\BibitemShut {NoStop}%
\bibitem [{\citenamefont {Thompson}\ \emph {et~al.}(2001)\citenamefont
  {Thompson}, \citenamefont {Edwards}, \citenamefont {McIntosh},\ and\
  \citenamefont {Townsend}}]{streams7}%
  \BibitemOpen
  \bibfield  {author} {\bibinfo {author} {\bibfnamefont {R.~M.}\ \bibnamefont
  {Thompson}}, \bibinfo {author} {\bibfnamefont {E.~D.}\ \bibnamefont
  {Edwards}}, \bibinfo {author} {\bibfnamefont {A.~R.}\ \bibnamefont
  {McIntosh}}, \ and\ \bibinfo {author} {\bibfnamefont {C.~R.}\ \bibnamefont
  {Townsend}},\ }\bibfield  {title} {\enquote {\bibinfo {title} {Allocation of
  effort in stream food-web studies: the best compromise?}}\ }\href@noop {}
  {\bibfield  {journal} {\bibinfo  {journal} {Marine and Freshwater Research}\
  }\textbf {\bibinfo {volume} {52}},\ \bibinfo {pages} {339–345} (\bibinfo
  {year} {2001})}\BibitemShut {NoStop}%
\bibitem [{\citenamefont {Ulanowicz}, \citenamefont {Bondavalli},\ and\
  \citenamefont {Egnotovich.}()}]{south_florida98}%
  \BibitemOpen
  \bibfield  {author} {\bibinfo {author} {\bibfnamefont {R.~E.}\ \bibnamefont
  {Ulanowicz}}, \bibinfo {author} {\bibfnamefont {C.}~\bibnamefont
  {Bondavalli}}, \ and\ \bibinfo {author} {\bibfnamefont {M.}~\bibnamefont
  {Egnotovich.}},\ }\bibfield  {title} {\enquote {\bibinfo {title} {Spatial and
  temporal variation in the structure of a freshwater food web},}\ }\href@noop
  {} {\bibinfo  {journal} {Network Analysis of Trophic Dynamics in South
  Florida Ecosystem, FY 97: The Florida Bay Ecosystem.}\ }\BibitemShut
  {NoStop}%
\bibitem [{\citenamefont {Field}\ \emph {et~al.}(1991)\citenamefont {Field},
  \citenamefont {Crawford}, \citenamefont {Wickens}, \citenamefont {Moloney},
  \citenamefont {Cochrane},\ and\ \citenamefont
  {Villacast\'in-Herrero}}]{Benguela_1}%
  \BibitemOpen
\bibfield  {journal} {  }\bibfield  {author} {\bibinfo {author} {\bibfnamefont
  {J.~G.}\ \bibnamefont {Field}}, \bibinfo {author} {\bibfnamefont {R.~J.~M.}\
  \bibnamefont {Crawford}}, \bibinfo {author} {\bibfnamefont {P.~A.}\
  \bibnamefont {Wickens}}, \bibinfo {author} {\bibfnamefont {C.~L.}\
  \bibnamefont {Moloney}}, \bibinfo {author} {\bibfnamefont {K.~L.}\
  \bibnamefont {Cochrane}}, \ and\ \bibinfo {author} {\bibfnamefont {C.~A.}\
  \bibnamefont {Villacast\'in-Herrero}},\ }\href@noop {} {\emph {\bibinfo
  {title} {Network analysis of {B}enguela pelagic food webs}}}\ (\bibinfo
  {publisher} {{B}enguela {E}cology {P}rogramme, {W}orkshop on {S}eal-{F}ishery
  {B}iological {I}nteractions. {U}niversity of {C}ape {T}own, 16-20, September,
  BEP/SW91/M5a},\ \bibinfo {address} {University of Cape Town},\ \bibinfo
  {year} {1991})\BibitemShut {NoStop}%
\bibitem [{\citenamefont {Yodzis}(1998)}]{Benguela_2}%
  \BibitemOpen
  \bibfield  {author} {\bibinfo {author} {\bibfnamefont {P.}~\bibnamefont
  {Yodzis}},\ }\bibfield  {title} {\enquote {\bibinfo {title} {Local
  trophodynamics and the interaction of marine mammals and fisheries in the
  {B}enguela ecosystem},}\ }\href@noop {} {\bibfield  {journal} {\bibinfo
  {journal} {J. Anim. Ecol.}\ }\textbf {\bibinfo {volume} {67}},\ \bibinfo
  {pages} {635--658} (\bibinfo {year} {1998})}\BibitemShut {NoStop}%
\bibitem [{\citenamefont {Havens}(1992)}]{bridge}%
  \BibitemOpen
  \bibfield  {author} {\bibinfo {author} {\bibfnamefont {K.}~\bibnamefont
  {Havens}},\ }\bibfield  {title} {\enquote {\bibinfo {title} {Scale and
  structure in natural food webs},}\ }\href@noop {} {\bibfield  {journal}
  {\bibinfo  {journal} {Science}\ }\textbf {\bibinfo {volume} {257}},\ \bibinfo
  {pages} {1107--1109} (\bibinfo {year} {1992})}\BibitemShut {NoStop}%
\bibitem [{\citenamefont {Memmott}, \citenamefont {Martinez},\ and\
  \citenamefont {Cohen}(2000)}]{Broom}%
  \BibitemOpen
  \bibfield  {author} {\bibinfo {author} {\bibfnamefont {J.}~\bibnamefont
  {Memmott}}, \bibinfo {author} {\bibfnamefont {N.~D.}\ \bibnamefont
  {Martinez}}, \ and\ \bibinfo {author} {\bibfnamefont {J.~E.}\ \bibnamefont
  {Cohen}},\ }\bibfield  {title} {\enquote {\bibinfo {title} {Predators,
  parasitoids and pathogens: species richness, trophic generality and body
  sizes in a natural food web},}\ }\href@noop {} {\bibfield  {journal}
  {\bibinfo  {journal} {J. Anim. Ecol.}\ }\textbf {\bibinfo {volume} {69}},\
  \bibinfo {pages} {1--15} (\bibinfo {year} {2000})}\BibitemShut {NoStop}%
\bibitem [{\citenamefont {Townsend}\ \emph {et~al.}(1998)\citenamefont
  {Townsend}, \citenamefont {Thompson}, \citenamefont {Mc{I}ntosh},
  \citenamefont {Kilroy}, \citenamefont {Edwards},\ and\ \citenamefont
  {Scarsbrook}}]{Canton_Stony}%
  \BibitemOpen
  \bibfield  {author} {\bibinfo {author} {\bibfnamefont {C.~R.}\ \bibnamefont
  {Townsend}}, \bibinfo {author} {\bibfnamefont {R.~M.}\ \bibnamefont
  {Thompson}}, \bibinfo {author} {\bibfnamefont {A.~R.}\ \bibnamefont
  {Mc{I}ntosh}}, \bibinfo {author} {\bibfnamefont {C.}~\bibnamefont {Kilroy}},
  \bibinfo {author} {\bibfnamefont {E.}~\bibnamefont {Edwards}}, \ and\
  \bibinfo {author} {\bibfnamefont {M.~R.}\ \bibnamefont {Scarsbrook}},\
  }\bibfield  {title} {\enquote {\bibinfo {title} {Disturbance, resource
  supply, and food-web architecture in streams},}\ }\href@noop {} {\bibfield
  {journal} {\bibinfo  {journal} {Ecol. Let.}\ }\textbf {\bibinfo {volume}
  {1}},\ \bibinfo {pages} {200--209} (\bibinfo {year} {1998})}\BibitemShut
  {NoStop}%
\bibitem [{\citenamefont {Bascompte}, \citenamefont {Meli\'an},\ and\
  \citenamefont {Sala}(2005)}]{Caribbean_2005}%
  \BibitemOpen
  \bibfield  {author} {\bibinfo {author} {\bibfnamefont {J.}~\bibnamefont
  {Bascompte}}, \bibinfo {author} {\bibfnamefont {C.}~\bibnamefont {Meli\'an}},
  \ and\ \bibinfo {author} {\bibfnamefont {E.}~\bibnamefont {Sala}},\
  }\bibfield  {title} {\enquote {\bibinfo {title} {Interaction strength
  combinations and the overfishing of a marine food web},}\ }\href@noop {} {\
  \textbf {\bibinfo {volume} {102}},\ \bibinfo {pages} {5443--5447} (\bibinfo
  {year} {2005})}\BibitemShut {NoStop}%
\bibitem [{\citenamefont {Lafferty}\ \emph {et~al.}(2006)\citenamefont
  {Lafferty}, \citenamefont {Hechinger}, \citenamefont {Shaw}, \citenamefont
  {Whitney},\ and\ \citenamefont {Kuris}}]{Carpinteria}%
  \BibitemOpen
  \bibfield  {author} {\bibinfo {author} {\bibfnamefont {K.~D.}\ \bibnamefont
  {Lafferty}}, \bibinfo {author} {\bibfnamefont {R.~F.}\ \bibnamefont
  {Hechinger}}, \bibinfo {author} {\bibfnamefont {J.~C.}\ \bibnamefont {Shaw}},
  \bibinfo {author} {\bibfnamefont {K.~L.}\ \bibnamefont {Whitney}}, \ and\
  \bibinfo {author} {\bibfnamefont {A.~M.}\ \bibnamefont {Kuris}},\ }\bibfield
  {title} {\enquote {\bibinfo {title} {Food webs and parasites in a salt marsh
  ecosystem},}\ }\href@noop {} {\bibfield  {journal} {\bibinfo  {journal}
  {Disease ecology: community structure and pathogen dynamics (ed. S. Collinge
  \& C. Ray)}\ ,\ \bibinfo {pages} {119--134}} (\bibinfo {year}
  {2006})}\BibitemShut {NoStop}%
\bibitem [{\citenamefont {Ulanowicz}(1986)}]{crystalD}%
  \BibitemOpen
  \bibfield  {author} {\bibinfo {author} {\bibfnamefont {R.~E.}\ \bibnamefont
  {Ulanowicz}},\ }\bibfield  {title} {\enquote {\bibinfo {title} {Growth and
  development: Ecosystems phenomenology. springer, new york. pp 69-79.}}\
  }\href@noop {} {\bibfield  {journal} {\bibinfo  {journal} {Network Analysis
  of Trophic Dynamics in South Florida Ecosystem, FY 97: The Florida Bay
  Ecosystem.}\ } (\bibinfo {year} {1986})}\BibitemShut {NoStop}%
\bibitem [{\citenamefont {Waide}\ and\ \citenamefont
  {Reagan}(1996)}]{El_verde}%
  \BibitemOpen
  \bibfield  {author} {\bibinfo {author} {\bibfnamefont {R.~B.}\ \bibnamefont
  {Waide}}\ and\ \bibinfo {author} {\bibfnamefont {W.~B.}\ \bibnamefont
  {Reagan}},\ }\href@noop {} {\emph {\bibinfo {title} {The Food Web of a
  Tropical Rainforest}}}\ (\bibinfo  {publisher} {University of Chicago
  Press},\ \bibinfo {address} {Chicago},\ \bibinfo {year} {1996})\BibitemShut
  {NoStop}%
\bibitem [{\citenamefont {Ulanowicz}, \citenamefont {Heymans},\ and\
  \citenamefont {Egnotovich}(2000)}]{everglades}%
  \BibitemOpen
  \bibfield  {author} {\bibinfo {author} {\bibfnamefont {R.~E.}\ \bibnamefont
  {Ulanowicz}}, \bibinfo {author} {\bibfnamefont {J.}~\bibnamefont {Heymans}},
  \ and\ \bibinfo {author} {\bibfnamefont {M.}~\bibnamefont {Egnotovich}},\
  }\bibfield  {title} {\enquote {\bibinfo {title} {Network analysis of trophic
  dynamics in south florida ecosystems},}\ }\href@noop {} {\bibfield  {journal}
  {\bibinfo  {journal} {Network Analysis of Trophic Dynamics in South Florida
  Ecosystems FY 99: The Graminoid Ecosystem.}\ } (\bibinfo {year}
  {2000})}\BibitemShut {NoStop}%
\bibitem [{\citenamefont {Martinez}\ \emph {et~al.}(1999)\citenamefont
  {Martinez}, \citenamefont {Hawkins}, \citenamefont {Dawah},\ and\
  \citenamefont {Feifarek}}]{Grass}%
  \BibitemOpen
  \bibfield  {author} {\bibinfo {author} {\bibfnamefont {N.~D.}\ \bibnamefont
  {Martinez}}, \bibinfo {author} {\bibfnamefont {B.~A.}\ \bibnamefont
  {Hawkins}}, \bibinfo {author} {\bibfnamefont {H.~A.}\ \bibnamefont {Dawah}},
  \ and\ \bibinfo {author} {\bibfnamefont {B.~P.}\ \bibnamefont {Feifarek}},\
  }\bibfield  {title} {\enquote {\bibinfo {title} {Effects of sampling effort
  on characterization of food-web structure},}\ }\href@noop {} {\bibfield
  {journal} {\bibinfo  {journal} {Ecology}\ }\textbf {\bibinfo {volume} {80}},\
  \bibinfo {pages} {1044–1055} (\bibinfo {year} {1999})}\BibitemShut
  {NoStop}%
\bibitem [{\citenamefont {Martinez}(1991)}]{Little_Rock}%
  \BibitemOpen
  \bibfield  {author} {\bibinfo {author} {\bibfnamefont {N.~D.}\ \bibnamefont
  {Martinez}},\ }\bibfield  {title} {\enquote {\bibinfo {title} {Artifacts or
  attributes? {E}ffects of resolution on the {L}ittle {R}ock {L}ake food
  web},}\ }\href@noop {} {\bibfield  {journal} {\bibinfo  {journal} {Ecol.
  Monogr.}\ }\textbf {\bibinfo {volume} {61}},\ \bibinfo {pages} {367--392}
  (\bibinfo {year} {1991})}\BibitemShut {NoStop}%
\bibitem [{\citenamefont {Riede}\ \emph {et~al.}(2011)\citenamefont {Riede},
  \citenamefont {Brose}, \citenamefont {Ebenman}, \citenamefont {Jacob},
  \citenamefont {Thompson}, \citenamefont {Townsend},\ and\ \citenamefont
  {Jonsson}}]{lough_hyne_1}%
  \BibitemOpen
  \bibfield  {author} {\bibinfo {author} {\bibfnamefont {J.}~\bibnamefont
  {Riede}}, \bibinfo {author} {\bibfnamefont {U.}~\bibnamefont {Brose}},
  \bibinfo {author} {\bibfnamefont {B.}~\bibnamefont {Ebenman}}, \bibinfo
  {author} {\bibfnamefont {U.}~\bibnamefont {Jacob}}, \bibinfo {author}
  {\bibfnamefont {R.}~\bibnamefont {Thompson}}, \bibinfo {author}
  {\bibfnamefont {C.}~\bibnamefont {Townsend}}, \ and\ \bibinfo {author}
  {\bibfnamefont {T.}~\bibnamefont {Jonsson}},\ }\bibfield  {title} {\enquote
  {\bibinfo {title} {Stepping in {E}lton's footprints: a general scaling model
  for body masses and trophic levels across ecosystems},}\ }\href@noop {}
  {\bibfield  {journal} {\bibinfo  {journal} {Ecology Letters}\ }\textbf
  {\bibinfo {volume} {14}},\ \bibinfo {pages} {169--178} (\bibinfo {year}
  {2011})}\BibitemShut {NoStop}%
\bibitem [{\citenamefont {Ekl\"of}\ \emph {et~al.}(2013)\citenamefont
  {Ekl\"of}, \citenamefont {Jacob}, \citenamefont {Kopp}, \citenamefont
  {Bosch}, \citenamefont {Castro-Urgal}, \citenamefont {Dalsgaard},
  \citenamefont {Chacoff}, \citenamefont {deSassi}, \citenamefont {Galetti},
  \citenamefont {Guimaraes}, \citenamefont {Lomáscolo}, \citenamefont
  {Martín~González}, \citenamefont {Pizo}, \citenamefont {Rader},
  \citenamefont {Rodrigo}, \citenamefont {Tylianakis}, \citenamefont
  {Vazquez},\ and\ \citenamefont {Allesina}}]{lough_hyne_2}%
  \BibitemOpen
  \bibfield  {author} {\bibinfo {author} {\bibfnamefont {A.}~\bibnamefont
  {Ekl\"of}}, \bibinfo {author} {\bibfnamefont {U.}~\bibnamefont {Jacob}},
  \bibinfo {author} {\bibfnamefont {J.}~\bibnamefont {Kopp}}, \bibinfo {author}
  {\bibfnamefont {J.}~\bibnamefont {Bosch}}, \bibinfo {author} {\bibfnamefont
  {R.}~\bibnamefont {Castro-Urgal}}, \bibinfo {author} {\bibfnamefont
  {B.}~\bibnamefont {Dalsgaard}}, \bibinfo {author} {\bibfnamefont
  {N.}~\bibnamefont {Chacoff}}, \bibinfo {author} {\bibfnamefont
  {C.}~\bibnamefont {deSassi}}, \bibinfo {author} {\bibfnamefont
  {M.}~\bibnamefont {Galetti}}, \bibinfo {author} {\bibfnamefont
  {P.}~\bibnamefont {Guimaraes}}, \bibinfo {author} {\bibfnamefont
  {S.}~\bibnamefont {Lomáscolo}}, \bibinfo {author} {\bibfnamefont
  {A.}~\bibnamefont {Martín~González}}, \bibinfo {author} {\bibfnamefont
  {M.}~\bibnamefont {Pizo}}, \bibinfo {author} {\bibfnamefont {R.}~\bibnamefont
  {Rader}}, \bibinfo {author} {\bibfnamefont {A.}~\bibnamefont {Rodrigo}},
  \bibinfo {author} {\bibfnamefont {J.}~\bibnamefont {Tylianakis}}, \bibinfo
  {author} {\bibfnamefont {D.}~\bibnamefont {Vazquez}}, \ and\ \bibinfo
  {author} {\bibfnamefont {S.}~\bibnamefont {Allesina}},\ }\bibfield  {title}
  {\enquote {\bibinfo {title} {The dimensionality of ecological networks},}\
  }\href@noop {} {\bibfield  {journal} {\bibinfo  {journal} {Ecology Letters}\
  }\textbf {\bibinfo {volume} {16}},\ \bibinfo {pages} {577--583} (\bibinfo
  {year} {2013})}\BibitemShut {NoStop}%
\bibitem [{\citenamefont {Ulanowicz}, \citenamefont {Bondavalli},\ and\
  \citenamefont {Egnotovich.}(1998)}]{mangrove}%
  \BibitemOpen
  \bibfield  {author} {\bibinfo {author} {\bibfnamefont {R.~E.}\ \bibnamefont
  {Ulanowicz}}, \bibinfo {author} {\bibfnamefont {C.}~\bibnamefont
  {Bondavalli}}, \ and\ \bibinfo {author} {\bibfnamefont {M.}~\bibnamefont
  {Egnotovich.}},\ }\bibfield  {title} {\enquote {\bibinfo {title} {Spatial and
  temporal variation in the structure of a freshwater food web},}\ }\href@noop
  {} {\bibfield  {journal} {\bibinfo  {journal} {Network Analysis of Trophic
  Dynamics in South Florida Ecosystem, FY 97: The Florida Bay Ecosystem.}\ }
  (\bibinfo {year} {1998})}\BibitemShut {NoStop}%
\bibitem [{\citenamefont {Mason}(2003)}]{michigan}%
  \BibitemOpen
  \bibfield  {author} {\bibinfo {author} {\bibfnamefont {D.}~\bibnamefont
  {Mason}},\ }\bibfield  {title} {\enquote {\bibinfo {title} {Quantifying the
  impact of exotic invertebrate invaders on food web structure and function in
  the great lakes: A network analysis approach},}\ }\href@noop {} {\bibfield
  {journal} {\bibinfo  {journal} {Interim Progress Report to the Great Lakes
  Fisheries Commission- yr 1}\ } (\bibinfo {year} {2003})}\BibitemShut
  {NoStop}%
\bibitem [{\citenamefont {Monaco}\ and\ \citenamefont
  {Ulanowicz}(1997)}]{narragan}%
  \BibitemOpen
  \bibfield  {author} {\bibinfo {author} {\bibfnamefont {M.~E.}\ \bibnamefont
  {Monaco}}\ and\ \bibinfo {author} {\bibfnamefont {R.~E.}\ \bibnamefont
  {Ulanowicz}},\ }\bibfield  {title} {\enquote {\bibinfo {title} {Comparative
  ecosystem trophic structure of three u.s mid-atlantic estuaries},}\
  }\href@noop {} {\bibfield  {journal} {\bibinfo  {journal} {Marine Ecology
  Progress Series}\ }\textbf {\bibinfo {volume} {161}},\ \bibinfo {pages}
  {239--254} (\bibinfo {year} {1997})}\BibitemShut {NoStop}%
\bibitem [{\citenamefont {Opitz}(1996)}]{Reef}%
  \BibitemOpen
  \bibfield  {author} {\bibinfo {author} {\bibfnamefont {S.}~\bibnamefont
  {Opitz}},\ }\bibfield  {title} {\enquote {\bibinfo {title} {Trophic
  interactions in {C}aribbean coral reefs},}\ }\href@noop {} {\bibfield
  {journal} {\bibinfo  {journal} {ICLARM Tech. Rep.}\ }\textbf {\bibinfo
  {volume} {43}},\ \bibinfo {pages} {341} (\bibinfo {year} {1996})}\BibitemShut
  {NoStop}%
\bibitem [{\citenamefont {Link}(2002)}]{Shelf}%
  \BibitemOpen
  \bibfield  {author} {\bibinfo {author} {\bibfnamefont {J.}~\bibnamefont
  {Link}},\ }\bibfield  {title} {\enquote {\bibinfo {title} {Does food web
  theory work for marine ecosystems?}}\ }\href@noop {} {\bibfield  {journal}
  {\bibinfo  {journal} {Mar. Ecol. Prog. Ser.}\ }\textbf {\bibinfo {volume}
  {230}},\ \bibinfo {pages} {1--9} (\bibinfo {year} {2002})}\BibitemShut
  {NoStop}%
\bibitem [{\citenamefont {Warren}(1989)}]{Skipwith}%
  \BibitemOpen
  \bibfield  {author} {\bibinfo {author} {\bibfnamefont {P.~H.}\ \bibnamefont
  {Warren}},\ }\bibfield  {title} {\enquote {\bibinfo {title} {Spatial and
  temporal variation in the structure of a freshwater food web},}\ }\href@noop
  {} {\bibfield  {journal} {\bibinfo  {journal} {Oikos}\ }\textbf {\bibinfo
  {volume} {55}},\ \bibinfo {pages} {299--311} (\bibinfo {year}
  {1989})}\BibitemShut {NoStop}%
\bibitem [{\citenamefont {Christian}\ and\ \citenamefont
  {Luczkovich}(1999)}]{St_Marks}%
  \BibitemOpen
  \bibfield  {author} {\bibinfo {author} {\bibfnamefont {R.~R.}\ \bibnamefont
  {Christian}}\ and\ \bibinfo {author} {\bibfnamefont {J.~J.}\ \bibnamefont
  {Luczkovich}},\ }\bibfield  {title} {\enquote {\bibinfo {title} {Organizing
  and understanding a winter's {S}eagrass foodweb network through effective
  trophic levels},}\ }\href@noop {} {\bibfield  {journal} {\bibinfo  {journal}
  {Ecol. Model.}\ }\textbf {\bibinfo {volume} {117}},\ \bibinfo {pages}
  {99--124} (\bibinfo {year} {1999})}\BibitemShut {NoStop}%
\bibitem [{\citenamefont {Goldwasser}\ and\ \citenamefont
  {Roughgarden}(1993)}]{St_Martin}%
  \BibitemOpen
  \bibfield  {author} {\bibinfo {author} {\bibfnamefont {L.}~\bibnamefont
  {Goldwasser}}\ and\ \bibinfo {author} {\bibfnamefont {J.~A.}\ \bibnamefont
  {Roughgarden}},\ }\bibfield  {title} {\enquote {\bibinfo {title}
  {Construction of a large {C}aribbean food web},}\ }\href@noop {} {\bibfield
  {journal} {\bibinfo  {journal} {Ecology}\ }\textbf {\bibinfo {volume} {74}},\
  \bibinfo {pages} {1216--1233} (\bibinfo {year} {1993})}\BibitemShut {NoStop}%
\bibitem [{\citenamefont {Jacob}\ \emph {et~al.}(2011)\citenamefont {Jacob},
  \citenamefont {Thierry}, \citenamefont {Brose}, \citenamefont {Arntz},
  \citenamefont {Berg}, \citenamefont {Brey}, \citenamefont {Fetzer},
  \citenamefont {Jonsson}, \citenamefont {Mintenbeck}, \citenamefont
  {Möllmann}, \citenamefont {Petchey}, \citenamefont {Riede},\ and\
  \citenamefont {Dunne}}]{weddell_sea}%
  \BibitemOpen
  \bibfield  {author} {\bibinfo {author} {\bibfnamefont {U.}~\bibnamefont
  {Jacob}}, \bibinfo {author} {\bibfnamefont {A.}~\bibnamefont {Thierry}},
  \bibinfo {author} {\bibfnamefont {U.}~\bibnamefont {Brose}}, \bibinfo
  {author} {\bibfnamefont {W.}~\bibnamefont {Arntz}}, \bibinfo {author}
  {\bibfnamefont {S.}~\bibnamefont {Berg}}, \bibinfo {author} {\bibfnamefont
  {T.}~\bibnamefont {Brey}}, \bibinfo {author} {\bibfnamefont {I.}~\bibnamefont
  {Fetzer}}, \bibinfo {author} {\bibfnamefont {T.}~\bibnamefont {Jonsson}},
  \bibinfo {author} {\bibfnamefont {K.}~\bibnamefont {Mintenbeck}}, \bibinfo
  {author} {\bibfnamefont {C.}~\bibnamefont {Möllmann}}, \bibinfo {author}
  {\bibfnamefont {O.}~\bibnamefont {Petchey}}, \bibinfo {author} {\bibfnamefont
  {J.}~\bibnamefont {Riede}}, \ and\ \bibinfo {author} {\bibfnamefont
  {J.}~\bibnamefont {Dunne}},\ }\bibfield  {title} {\enquote {\bibinfo {title}
  {The role of body size in complex food webs},}\ }\href@noop {} {\bibfield
  {journal} {\bibinfo  {journal} {Advances in Ecological Research}\ }\textbf
  {\bibinfo {volume} {45}},\ \bibinfo {pages} {181--223} (\bibinfo {year}
  {2011})}\BibitemShut {NoStop}%
\bibitem [{\citenamefont {Hall}\ and\ \citenamefont {Raffaelli}(1991)}]{Ythan}%
  \BibitemOpen
  \bibfield  {author} {\bibinfo {author} {\bibfnamefont {S.~J.}\ \bibnamefont
  {Hall}}\ and\ \bibinfo {author} {\bibfnamefont {D.}~\bibnamefont
  {Raffaelli}},\ }\bibfield  {title} {\enquote {\bibinfo {title} {Food-web
  patterns: lessons from a species-rich web},}\ }\href@noop {} {\bibfield
  {journal} {\bibinfo  {journal} {J. Anim. Ecol.}\ }\textbf {\bibinfo {volume}
  {60}},\ \bibinfo {pages} {823--842} (\bibinfo {year} {1991})}\BibitemShut
  {NoStop}%
\bibitem [{\citenamefont {Duch}\ and\ \citenamefont
  {Arenas}(2005)}]{CElegans_metabolic}%
  \BibitemOpen
  \bibfield  {author} {\bibinfo {author} {\bibfnamefont {J.}~\bibnamefont
  {Duch}}\ and\ \bibinfo {author} {\bibfnamefont {A.}~\bibnamefont {Arenas}},\
  }\bibfield  {title} {\enquote {\bibinfo {title} {Community identification
  using extremal optimization},}\ }\href@noop {} {\bibfield  {journal}
  {\bibinfo  {journal} {Physical Review E}\ }\textbf {\bibinfo {volume} {72}},\
  \bibinfo {pages} {027104} (\bibinfo {year} {2005})}\BibitemShut {NoStop}%
\bibitem [{\citenamefont {Sanz}\ \emph {et~al.}(2011)\citenamefont {Sanz},
  \citenamefont {Navarro}, \citenamefont {Arbués}, \citenamefont {Martín},
  \citenamefont {Marijuán},\ and\ \citenamefont {Moreno}}]{EColi_bmc}%
  \BibitemOpen
  \bibfield  {author} {\bibinfo {author} {\bibfnamefont {J.}~\bibnamefont
  {Sanz}}, \bibinfo {author} {\bibfnamefont {J.}~\bibnamefont {Navarro}},
  \bibinfo {author} {\bibfnamefont {A.}~\bibnamefont {Arbués}}, \bibinfo
  {author} {\bibfnamefont {C.}~\bibnamefont {Martín}}, \bibinfo {author}
  {\bibfnamefont {P.~C.}\ \bibnamefont {Marijuán}}, \ and\ \bibinfo {author}
  {\bibfnamefont {Y.}~\bibnamefont {Moreno}},\ }\bibfield  {title} {\enquote
  {\bibinfo {title} {The transcriptional regulatory network of mycobacterium
  tuberculosis},}\ }\href@noop {} {\bibfield  {journal} {\bibinfo  {journal}
  {PLoS ONE}\ }\textbf {\bibinfo {volume} {6}},\ \bibinfo {pages} {e22178}
  (\bibinfo {year} {2011})}\BibitemShut {NoStop}%
\bibitem [{\citenamefont {Yu}\ and\ \citenamefont {Gerstein}()}]{Yeast_reg}%
  \BibitemOpen
  \bibfield  {author} {\bibinfo {author} {\bibfnamefont {H.}~\bibnamefont
  {Yu}}\ and\ \bibinfo {author} {\bibfnamefont {M.}~\bibnamefont {Gerstein}},\
  }\href@noop {} {\bibinfo  {journal} {Proc. Natl. Acad. Sci. USA}\
  }\BibitemShut {NoStop}%
\bibitem [{\citenamefont {Ma'ayan}\ \emph {et~al.}(2008)\citenamefont
  {Ma'ayan}, \citenamefont {Cecchi}, \citenamefont {Wagner}, \citenamefont
  {Raob}, \citenamefont {Iyengara},\ and\ \citenamefont
  {Stolovitzky}}]{Cecchi-PNAS}%
  \BibitemOpen
\bibfield  {journal} {  }\bibfield  {author} {\bibinfo {author} {\bibfnamefont
  {A.}~\bibnamefont {Ma'ayan}}, \bibinfo {author} {\bibfnamefont {G.~A.}\
  \bibnamefont {Cecchi}}, \bibinfo {author} {\bibfnamefont {J.}~\bibnamefont
  {Wagner}}, \bibinfo {author} {\bibfnamefont {A.~R.}\ \bibnamefont {Raob}},
  \bibinfo {author} {\bibfnamefont {R.}~\bibnamefont {Iyengara}}, \ and\
  \bibinfo {author} {\bibfnamefont {G.}~\bibnamefont {Stolovitzky}},\
  }\bibfield  {title} {\enquote {\bibinfo {title} {Ordered cyclic motifs
  contribute to dynamic stability in biological and engineered networks},}\
  }\href@noop {} {\bibfield  {journal} {\bibinfo  {journal} {PNAS}\ }\textbf
  {\bibinfo {volume} {105}},\ \bibinfo {pages} {19235--19240} (\bibinfo {year}
  {2008})}\BibitemShut {NoStop}%
\bibitem [{\citenamefont {Rodríguez-Caso}, \citenamefont {Corominas-Murtra},\
  and\ \citenamefont {Solé}(2009)}]{computation_GRN}%
  \BibitemOpen
  \bibfield  {author} {\bibinfo {author} {\bibfnamefont {C.}~\bibnamefont
  {Rodríguez-Caso}}, \bibinfo {author} {\bibfnamefont {B.}~\bibnamefont
  {Corominas-Murtra}}, \ and\ \bibinfo {author} {\bibfnamefont {R.~V.}\
  \bibnamefont {Solé}},\ }\bibfield  {title} {\enquote {\bibinfo {title} {On
  the basic computational structure of gene regulatory networks},}\ }\href@noop
  {} {\bibfield  {journal} {\bibinfo  {journal} {Mol. BioSyst.}\ }\textbf
  {\bibinfo {volume} {5}},\ \bibinfo {pages} {1617--1629} (\bibinfo {year}
  {2009})}\BibitemShut {NoStop}%
\bibitem [{\citenamefont {Watts}\ and\ \citenamefont
  {Strogatz}(1998)}]{CElegans_neural}%
  \BibitemOpen
  \bibfield  {author} {\bibinfo {author} {\bibfnamefont {D.~J.}\ \bibnamefont
  {Watts}}\ and\ \bibinfo {author} {\bibfnamefont {S.~H.}\ \bibnamefont
  {Strogatz}},\ }\bibfield  {title} {\enquote {\bibinfo {title} {Collective
  dynamics of 'small-world' networks},}\ }\href@noop {} {\bibfield  {journal}
  {\bibinfo  {journal} {Nature}\ }\textbf {\bibinfo {volume} {393}},\ \bibinfo
  {pages} {440--442} (\bibinfo {year} {1998})}\BibitemShut {NoStop}%
\end{thebibliography}

%

\clearpage

\onecolumngrid
\appendix

\LTcapwidth=\textwidth
\squeezetable

\begin{longtable}{ccccccccc}
Name & \text{S} & \text{$<k>$} & \text{Type} & \text{$\xi$ Z-score} & \text{}  & \text{$\eta$ Z-score} & \text{}  &\text{ref}  \\
\hline
Food webs & & & & & & & & \\
\hline
Akatore Stream & 85 & 2.67 & Stream & 2.650 & (**) & 685.250 & (***) & \cite{streams5,streams6,streams7} \\
Florida Bay (dry season) & 122 & 14.75 & Marine & 68.536 & (***) & 49.307 & (***) & \cite{south_florida98} \\
Florida Bay (wet season) & 122 & 14.48 & Marine & 55.779 & (***) & 49.175 & (***) & \cite{south_florida98} \\
Benguela Current & 29 & 7.00 & Marine & 11.492 & (***) & 7.028 & (***) & \cite{Benguela_1,Benguela_2} \\
Berwick Stream & 79 & 3.04 & Stream & 9.206 & (***) & 6.566 & (***) & \cite{streams5,streams6,streams7} \\
Blackrock Stream & 87 & 4.31 & Stream & 6.350 & (***) & 6.297 & (***) & \cite{streams5,streams6,streams7} \\
Bridge Brook Lake & 25 & 4.28 & Lake & 14.676 & (***) & 73.785 & (***) & \cite{bridge} \\
Broad Stream & 95 & 5.95 & Stream & 15.423 & (***) & 8.742 & (***) & \cite{streams5,streams6,streams7} \\
Scotch Broom & 85 & 2.62 & Terrestrial & 10.644 & (***) & 23.209 & (***) & \cite{Broom} \\
Canton Creek & 102 & 6.83 & Stream & 13.678 & (***) & 20.675 & (***) & \cite{Canton_Stony} \\
Caribbean (2005) & 249 & 13.31 & Marine & 32.603 & (***) & 54.878 & (***) & \cite{Caribbean_2005} \\
Carpinteria Salt Marsh Reserve & 128 & 4.23 & Marine & 24.123 & (***) & 23.272 & (***) & \cite{Carpinteria} \\
Catlins Stream & 49 & 2.24 & Stream & 4.751 & (***) & 0.238 & () & \cite{streams5,streams6,streams7} \\
Cayman Islands & 261 & 14.43 & Marine & 52.739 & (***) & 34.324 & (***) & \cite{Caribbean_2005} \\
Chesapeake Bay & 31 & 2.19 & Marine & 7.997 & (***) & 4.868 & (***) & \cite{chesapeake1,chesapeake2} \\
Crystal Lake (control) & 20 & 2.55 & Lake & 5.454 & (***) & 15.001 & (***) & \cite{crystalD} \\
Crystal Lake (delta) & 20 & 1.65 & Lake & 2.226 & (**) & 4.405 & (***) & \cite{crystalD} \\
Cuba & 261 & 14.84 & Marine & 61.826 & (***) & 28.068 & (***) & \cite{Caribbean_2005} \\
Cypress (dry season) & 65 & 6.89 & Terrestrial & 18.791 & (***) & 187.024 & (***) & \cite{south_florida98} \\
Cypress (wet season) & 65 & 6.75 & Terrestrial & 25.329 & (***) & 251.612 & (***) & \cite{south_florida98} \\
Dempsters Stream (autum) & 86 & 4.83 & Stream & 10.116 & (***) & 42.483 & (***) & \cite{streams5,streams6,streams7} \\
El Verde Rainforest & 155 & 9.74 & Terrestrial & 56.909 & (***) & 110.125 & (***) & \cite{El_verde} \\
Everglades Graminoid Marshes & 63 & 9.79 & Terrestrial & 51.266 & (***) & 16.429 & (***) & \cite{everglades} \\
Florida Bay & 122 & 14.48 & Marine & 55.670 & (***) & 276.139 & (***) & \cite{south_florida98} \\
Graminoid Marshes (dry) & 63 & 9.79 & Terrestrial & 51.266 & (***) & 16.429 & (***) & \cite{everglades} \\
Graminoid Marshes (wet) & 63 & 9.79 & Terrestrial & 51.266 & (***) & 16.429 & (***) & \cite{everglades} \\
Grassland & 61 & 1.59 & Terrestrial & 2.917 & (**) & 3.516 & (***) & \cite{Grass} \\
Healy Stream & 96 & 6.60 & Stream & 11.242 & (***) & 67.238 & (***) & \cite{streams5,streams6,streams7} \\
Jamaica & 263 & 15.61 & Marine & 82.760 & (***) & 179.738 & (***) & \cite{Caribbean_2005} \\
Kyeburn Stream & 98 & 6.42 & Stream & 8.142 & (***) & 65.368 & (***) & \cite{streams5,streams6,streams7} \\
Little Rock Lake & 92 & 10.84 & Lake & 54.022 & (***) & 215.878 & (***) & \cite{Little_Rock} \\
Lough Hyne & 349 & 14.66 & Marine & 67.464 & (***) & 645.370 & (***) & \cite{lough_hyne_1,lough_hyne_2} \\
Mangrove Estuary (dry season) & 91 & 12.63 & Marine & 36.641 & (***) & 21.644 & (***) & \cite{mangrove} \\
Mangrove Estuary (wet season) & 91 & 12.65 & Marine & 36.163 & (***) & 77.527 & (***) & \cite{mangrove} \\
Michigan Lake & 35 & 3.69 & Lake & 24.402 & (***) & 36.655 & (***) & \cite{michigan} \\
Mondego Estuary & 42 & 6.64 & Marine & 13.237 & (***) & 11.951 & (***) & \cite{mondego} \\
Narragansett Bay & 31 & 3.65 & Marine & 6.841 & (***) & 28.552 & (***) & \cite{narragan} \\
Caribbean Reef & 50 & 11.12 & Marine & 11.179 & (***) & 2862.490 & (***) & \cite{Reef} \\
N.E. Shelf & 79 & 17.76 & Marine & 24.212 & (***) & 27.151 & (***) & \cite{Shelf} \\
Skipwith Pond & 25 & 7.88 & Lake & 2.504 & (**) & 15.742 & (***) & \cite{Skipwith} \\
St. Marks Estuary & 48 & 4.60 & Marine & 21.388 & (***) & 22.591 & (***) & \cite{St_Marks} \\
St. Martin Island & 42 & 4.88 & Terrestrial & 12.730 & (***) & 17.923 & (***) & \cite{St_Martin} \\
Stony Stream & 109 & 7.61 & Stream & 13.525 & (***) & 10.515 & (***) & \cite{Canton_Stony} \\
Troy Stream & 78 & 2.32 & Stream & 5.686 & (***) & 6.622 & (***) & \cite{streams5,streams6,streams7} \\
Weddell Sea & 483 & 31.81 & Marine & 120.062 & (***) & 517.488 & (***) & \cite{weddell_sea} \\
Ythan Estuary & 82 & 4.82 & Marine & 7.713 & (***) & 21.531 & (***) & \cite{Ythan} \\
\hline
Biological & & & & & & & & \\
\hline
C. Elengans metabolic & 453 & 4.50 & TRN & 7.726 & (***) & 125.040 & (***) & \cite{CElegans_metabolic} \\
E. Coli transcription & 1037 & 2.59 & TRN & 51.538 & (***) & 78.798 & (***) & \cite{EColi_bmc} \\
Mus musculus transcription & 73 & 1.62 & TRN & 3.908 & (***) & 17.283 & (***) & \cite{Yeast_reg} \\
Mammalian signalling & 599 & 2.34 & Cell signalling & 21.686 & (***) & 43.677 & (***) & \cite{Cecchi-PNAS} \\
B. Subtilis transcription & 814 & 1.69 & TRN & 2.305 & (**) & 85.051 & (***) & \cite{computation_GRN} \\
M. Tuberculosis transcription & 1624 & 1.98 & TRN & -9.637 & (***) & 11.029 & (***) & \cite{EColi_bmc} \\
\hline
Other Networks & & & & & & & & \\
\hline
C. Elegans neural & 297 & 7.90 & neural & 22.429 & (***) & 45.026 & (***) & \cite{CElegans_neural} \\
Word adjacency & 50 & 2.02 & words & 6.479 & (***) & 9.589 & (***) &  \\
U.S.A airports & 1226 & 2.13 & airports & 49.869 & (***) & 56.691 &
(***) & \cite{Cecchi-PNAS} \\\\
\caption{List of empirical networks considered in this work. The first
  column shows their names, the fourth their types, and the last
  one references to the literature. The z-scores of intervalities
  $\xi$ and $\eta$ are computed with respect to a null model
  consisting in many randomisations of each network which preserve
  both in- and out-degree sequences.
  The number of ``stars'' after
  each z-score indicates whether the z-score is between $1$ and $2$ ($*$),
  $2$ and $3$ ($**$), or larger than $3$ ($***$).}
\label{table_zscore}
\end{longtable}

\end{document}